\documentclass{ieeeaccess}
\usepackage[numbers,sort&compress]{natbib}
\usepackage{epsf,graphicx}
\usepackage{amsmath,amssymb,amsfonts}
\usepackage{subfigure} 
\graphicspath{{./figs/}}
\usepackage{float}
\usepackage{multirow}
\usepackage{multicol}
\usepackage{cleveref}
\usepackage{textcomp}
\usepackage{algorithmic}

\usepackage{bm}
\makeatletter
\AtBeginDocument{\DeclareMathVersion{bold}
\SetSymbolFont{operators}{bold}{T1}{times}{b}{n}
\SetSymbolFont{NewLetters}{bold}{T1}{times}{b}{it}
\SetMathAlphabet{\mathrm}{bold}{T1}{times}{b}{n}
\SetMathAlphabet{\mathit}{bold}{T1}{times}{b}{it}
\SetMathAlphabet{\mathbf}{bold}{T1}{times}{b}{n}
\SetMathAlphabet{\mathtt}{bold}{OT1}{pcr}{b}{n}
\SetSymbolFont{symbols}{bold}{OMS}{cmsy}{b}{n}
\renewcommand\boldmath{\@nomath\boldmath\mathversion{bold}}}
\makeatother

\def\BibTeX{{\rm B\kern-.05em{\sc i\kern-.025em b}\kern-.08em
    T\kern-.1667em\lower.7ex\hbox{E}\kern-.125emX}}

\begin{document}
\history{Date of publication xxxx 00, 0000, date of current version xxxx 00, 0000.}
\doi{10.1109/ACCESS.2025.3567182}

\title{A Spin Photonic Topological Metasurface Based On A Kagome Lattice For Use In Leaky-wave Antenna Application}
\author{SAYYED AHMAD ABTAHI\authorrefmark{1},
        MOHSEN MADDAHALI\authorrefmark{1},
        AND AHMAD BAKHTAFROUZ\authorrefmark{1}}
\address[1]{Department of Electrical and Computer Engineering,
        Isfahan University of Technology,
        Isfahan 84156-83111, Iran }
\markboth
{S. A. Abtahi \headeretal: A Spin Photonic Topological Metasurface Based On A Kagome Lattice For Use In Leaky-wave Antenna Application}
{S. A. Abtahi \headeretal: A Spin Photonic Topological Metasurface Based On A Kagome Lattice For Use In Leaky-wave Antenna Application}
\corresp{Corresponding author: M. Maddahali (e-mail: maddahali@iut.ac.ir).}

\begin{abstract}
The emerging field of topological metasurfaces offers unique advantages, particularly in robustness against backscattering in low-profile structures. The lattice configuration of these structures significantly influences the ability to achieve sharp turns in the propagation path. One of the most studied lattices in condensed matter physics is the kagome lattice, characterized by its hexagonal Brillouin zone, which displays a Dirac cone in its dispersion diagram. 
Previous research on kagome lattices in photonic topological insulators has primarily focused on valley types of insulators. This article introduces a spin topological metasurface based on the kagome lattice and its unit cell, enabling a broad range of sharp turns and propagation paths. 
The unit cell is compared to its hexagonal and 60-degree rhombic counterparts, and a parametric study of its dimensions is conducted. As a result of this research, a new X-band leaky-wave antenna designed in the kagome lattice with an armchair arrangement interface has been developed. This antenna provides two forward and two backward beams, each pair achieving an approximately 50-degree scan within the 8.8 to 11.1 GHz bandwidth.
\end{abstract}


\titlepgskip=-21pt

\maketitle

\section{Introduction}

\PARstart{T}{opological} metasurfaces are emerging in this decade as part of the third generation of metasurfaces\cite{EM_Meta}. They exhibit unique characteristics, such as immunity to backscattering during sharp turns and resilience against manufacturing defects\cite{beginner}. These structures are part of photonic topological insulators (PTIs), classified into two groups based on whether time-reversal symmetry is preserved or broken. The first group is called Chern PTIs, which consist of photonic crystals made from ferrite rods arranged in a square\cite{ChernObs} or honeycomb lattice\cite{poo2011}. Time-reversal symmetry is broken by applying a static magnetic field. Conversely, the second group is distinctly categorized into two types of valley or spin PTIs. This classification is achieved through the breaking of inversion symmetry\cite{diaSPIE} and electromagnetic duality\cite{bisharat2019}, respectively, which facilitates propagation in two distinct opposing directions.\cite{ozawa}.

In the spin PTIs, the quantum spin Hall effect is emulated using electromagnetic duality, characterized by the condition (\(\overline{\overline{\epsilon}} = \overline{\overline{\mu}}\)), in spin-degenerate metamaterials. This setup can lead to either fourfold \cite{bisharat2019} or twofold \cite{davisPhDthesis} degeneracies at the K and K' symmetry points within the irreducible Brillouin zone. Consequently, by enforcing bi-anisotropy, this degeneracy is broken, resulting in the emergence of a non-trivial bandgap \cite{khanikaev2013}.

Spin topological metasurfaces are designed on honeycomb\cite{bisharat2019,diaSPIE} and 60-degree\cite{davisRhombic} and 30-degree\cite{abtahi2} rhombic lattices. It is important to note that although the waves propagating in the spin topological metasurfaces are line waves, these structures are distinct from dual impedance metasurfaces.\cite{xulinewave, salardual, khodadadidual,KhavasiLinewave}. The wave propagating in dual impedance metasurfaces lacks topological protection and will scatter at sharp bends.\cite{AdvMet}.    

In addition to guiding waves, these structures also have been used as leaky-wave antennas\cite{chernAntenna,ChernPLWA,SinghAntenna,abtahi2,valleyLWA}. In \cite{SinghAntenna}, the authors utilized spin topological metasurfaces based on a honeycomb lattice arranged in a zigzag pattern. However, the propagation through this zigzag configuration leads to coupling between the radiating elements, resulting in a non-practical radiation pattern. In \cite{abtahi2}, the coupling problem was addressed using a 30-degree rhombic lattice. However, this lattice structure has a gap in the edge modes, which are situated in the slow wave region. As a result, the 30-degree rhombic structure is not suitable for waveguiding applications.

This article presents a spin topological metasurface that utilizes a kagome lattice configuration. It begins by detailing the characteristics of the corresponding unit cell and observing the unidirectional propagation of each pseudospin, noting its robustness during sharp turns. Following this, a parametric study is conducted to determine optimal dimensions for operation in X-band frequencies. Drawing inspiration from the results of \cite{abtahiArmchair}, which focused on the armchair arrangement, the interface line is designed in an armchair configuration for use in a leaky-wave antenna application.

The proposed leaky-wave antenna is designed to radiate two beams forward and two beams backward simultaneously. It operates at a frequency range of 8.8 to 11.1 GHz, allowing for a 50-degree scan for backward beams and a 47-degree scan for forward beams. The simulations conducted using Ansys HFSS have been validated by repeating them in CST Studio.     
\section{kagome lattice metasurface characteristics}
The kagome lattice is interested in condensed matter physics for inclusion of Dirac cone, flat band and Van Hove singularity\cite{TopologicalKagomeWang,TopologicalKagomeYin}. The non-triviality of the lattice has been proven for electronic structures\cite{TopologicalKagomeGuo} and many valley topological photonic crystals has been proposed based on the lattice\cite{Valley2019Kagome,Valley2019Kagome2,Valley2019Kagome3,Valley2020Kagome,Valley2021Kagome,Valley2022Kagome,Valley2023Kagome,Valley2024Kagome}.

\begin{figure}[t!]
	\begin{center}
	\subfigure[]{
	\includegraphics[width=0.1\textwidth]{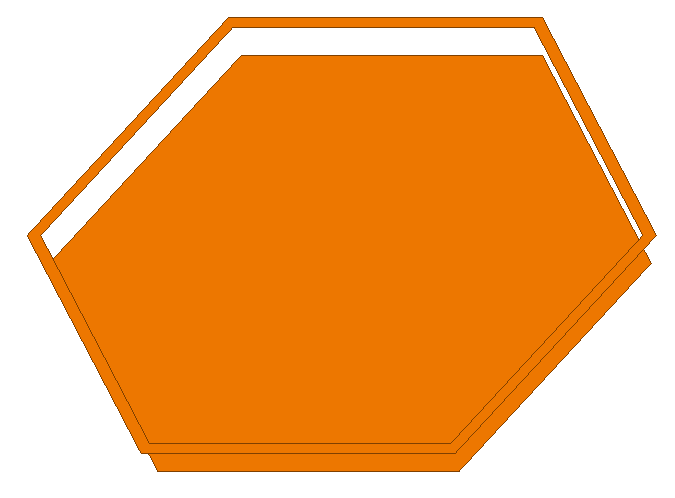}
	\label{Hex_cell}
	}
	\hspace{0.001cm}
	\subfigure[]{
	\includegraphics[width=0.14\textwidth]{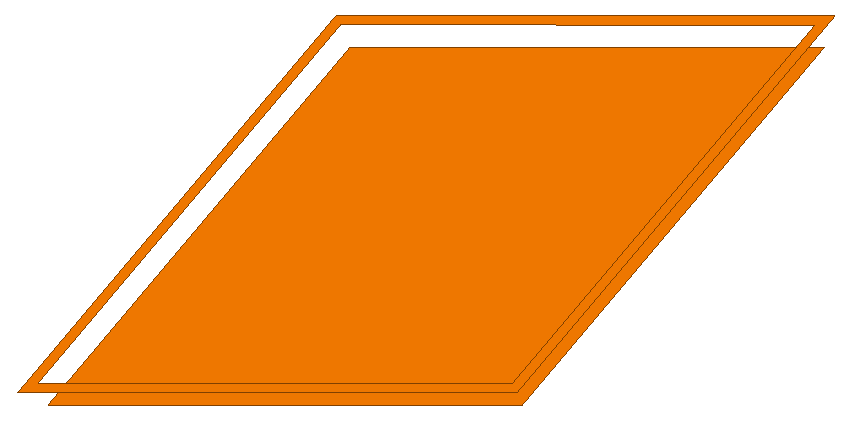}
	\label{Rhombic_60_2}
	}
	\hspace{0.001cm}
	\subfigure[]{
	\includegraphics[width=0.15\textwidth]{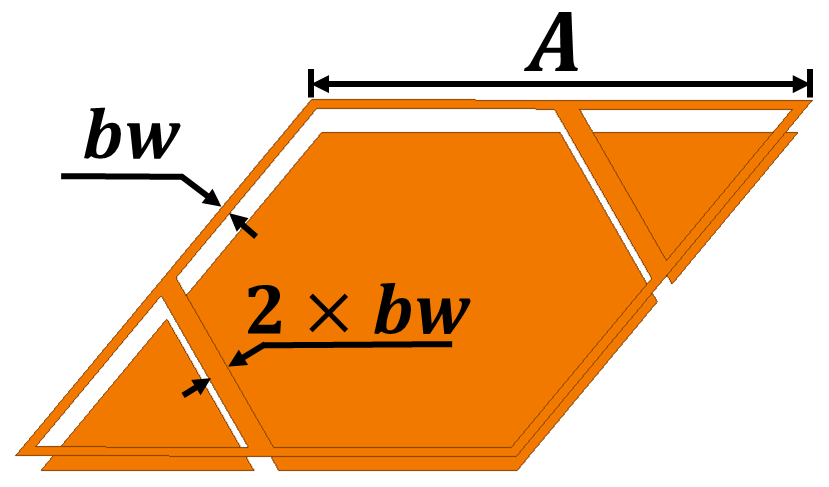}
	\label{Kagome_cell}
	}
	\end{center}
\caption{\subref{Hex_cell} The proposed hexagonal unit cell of the honeycomb lattice, which has been introduced in \cite{bisharat2019} in zigzag arrangement, \subref{Rhombic_60_2} the proposed 60-degree rhombic unit cell which has been introduced in \cite{davisRhombic} and \subref{Kagome_cell} the proposed kagome unit cell which is introduced in this paper. The period length and border width are marked by \textbf{A} and \textbf{bw}, respectively.}
\label{Cell}
\end{figure}

Despite mentioned hilarious efforts, there was no low profile spin PTI on Kagome lattice to the best of our knowledge. The method of using EM duality in complementary cell to achieve a low profile spin PTI introduced in \cite{bisharat2019}, for the first time, features a hexagonal unit cell shown in Fig.\ref{Hex_cell}. Inspired by the hexagonal Brillouin zone of a the hexagonal lattice, a 60-degree rhombic unit cell (Fig.\ref{Rhombic_60_2}) with a similar Brillouin zone has been introduced\cite{davisRhombic}, and its non-triviality has been proven\cite{davisPhDthesis}. The kagome lattice is another structure that exhibits a hexagonal Brillouin zone \cite{KagomeBrillouin}, which will be constructed using the presented unit cell shown in Fig.\ref{Kagome_cell}. 

In the following, the dispersion diagram of kagome cell calculated and compared with the 60-degree rhombic cell. Also, the dispersion diagram of one ribbon of each mentioned cell in Fig.\ref{Cell} determined to observe the edge mode presence and behavior. After that a parametric study on the dimensions of the kagome cell has been conducted , to find appropriate design of the structure in the X-band frequencies. 
\subsection{Dispersion diagram}

The period length and border width are set to 20 mm and 0.433 mm, respectively, to facilitate a direct comparison with the 60-degree rhombic unit cell described in \cite{davisRhombic}. Additionally, the substrate has been selected as Rogers/Duroid 5880, which has a thickness of 1.57 mm and a permittivity of 2.2.

\begin{figure}[t!]
	\begin{center}
	\subfigure[]{
	\raisebox{0.05\height}{\includegraphics[width=0.16\textwidth]{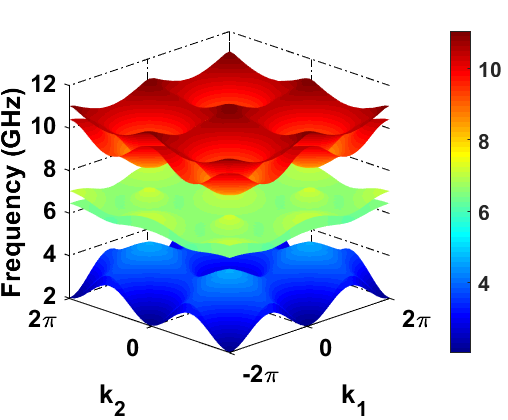}}
	\label{Rhombic_60_3D_dispersion_20_0433}
	}
	\hspace{0.00001cm}
	\subfigure[]{
	\includegraphics[width=0.28\textwidth]{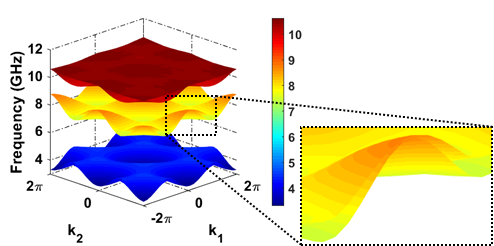}
	\label{Kagome_3D_dispersion_20_0433_tight}
	}
	\end{center}
\caption{ 
The diagrams in \subref{Rhombic_60_3D_dispersion_20_0433} and \subref{Kagome_3D_dispersion_20_0433_tight} illustrate the dispersion of a 60-degree rhombic and a kagome unit cell in 3D, respectively. The inset clearly indicates the degeneracy of the two bands.   
  }
\label{3D_dispersion}
\end{figure}

The 3D dispersion of the 60-degree rhombic and kagome unit cells is illustrated in Fig.\ref{Rhombic_60_3D_dispersion_20_0433} and \subref{Kagome_3D_dispersion_20_0433_tight}, which demonstrate the degeneracy of the bands and the presence of a complete bandgap between them. Notably, the first bandgap of the kagome unit cell is significantly wider than that of the 60-degree rhombic unit cell. 

\begin{figure}[h!]
	\begin{center}
	\subfigure[]{
	\raisebox{0.2\height}{\includegraphics[width=0.06\textwidth]{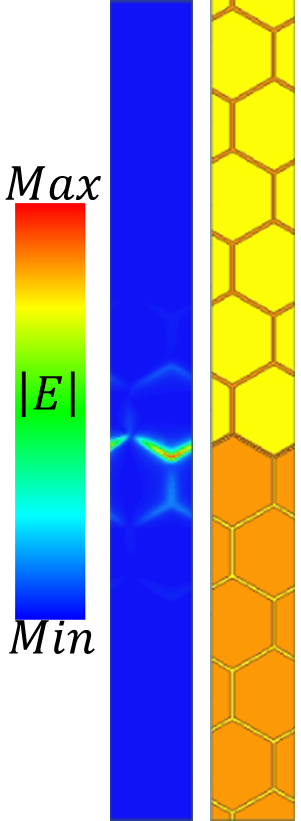}}
	\includegraphics[width=0.38\textwidth]{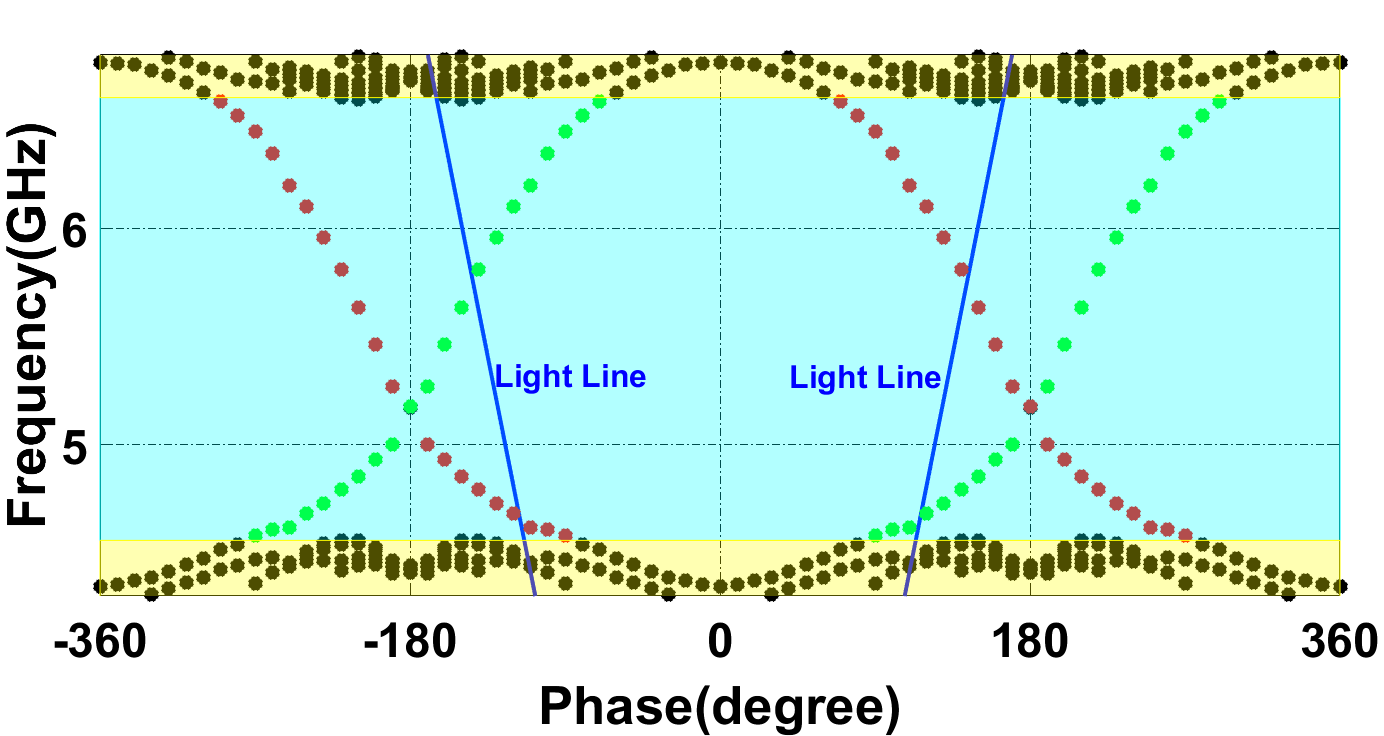}
	\label{ZigZag_Ribbon_dispersion_a_20_b_0433_SuperExtended}
	}
	\hspace{0.0001cm}
	\subfigure[]{
	\raisebox{0.2\height}{\includegraphics[width=0.055\textwidth]{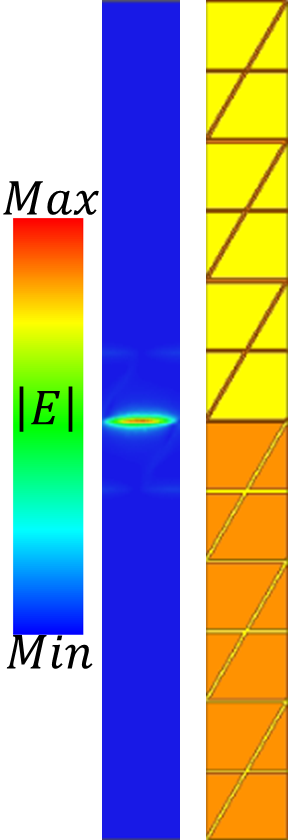}}
	\includegraphics[width=0.38\textwidth]{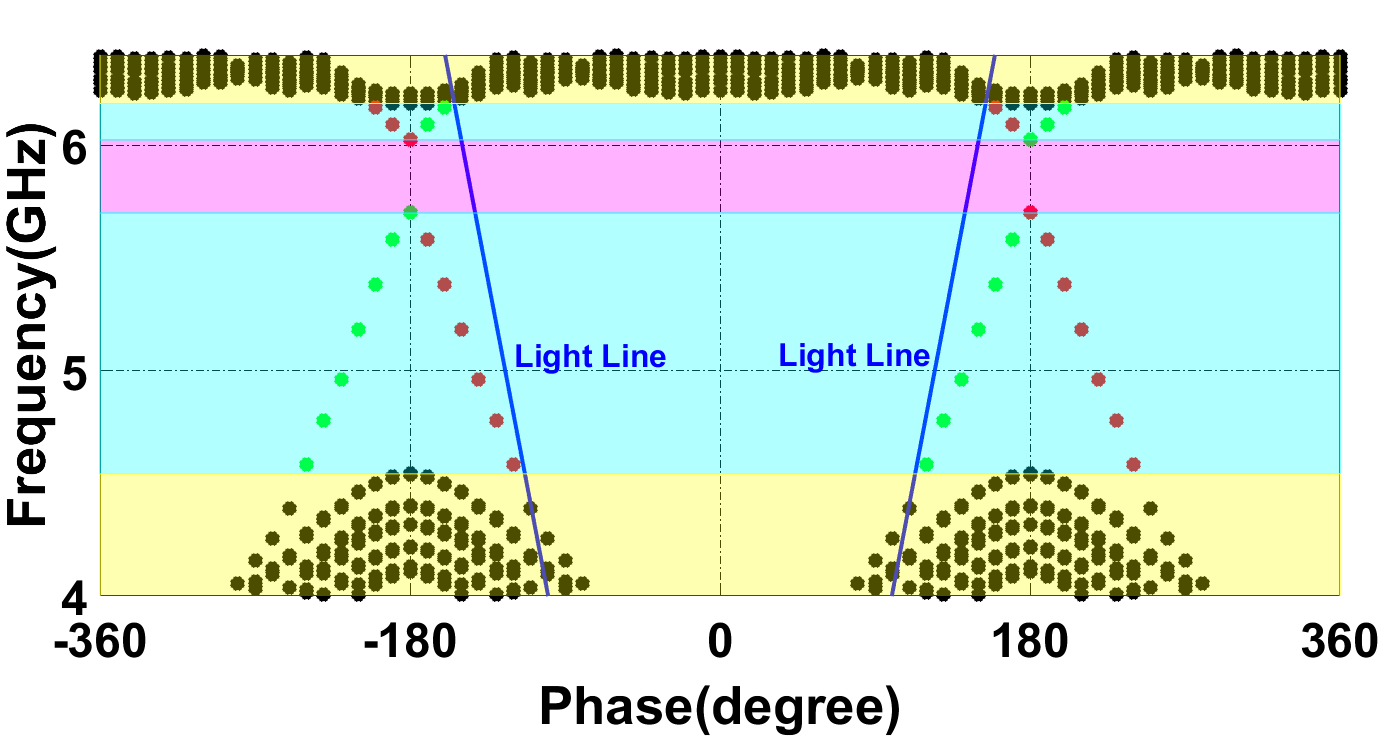}
	\label{Ribbon_Rhombic60_a_20_b_0433_SuperExtended}
	}
	\hspace{0.0001cm}
	\subfigure[]{
	\raisebox{0.2\height}{\includegraphics[width=0.058\textwidth]{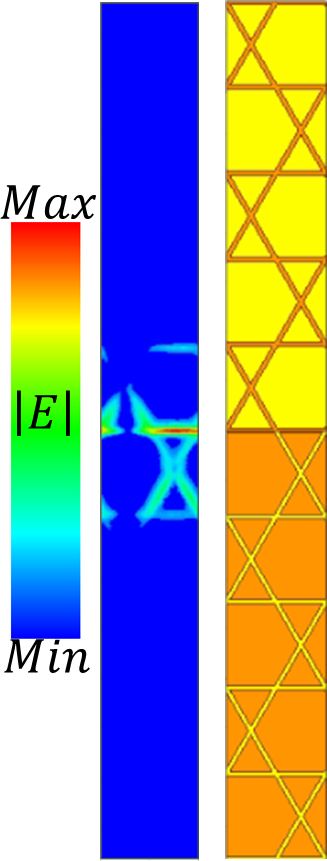}}
	\includegraphics[width=0.38\textwidth]{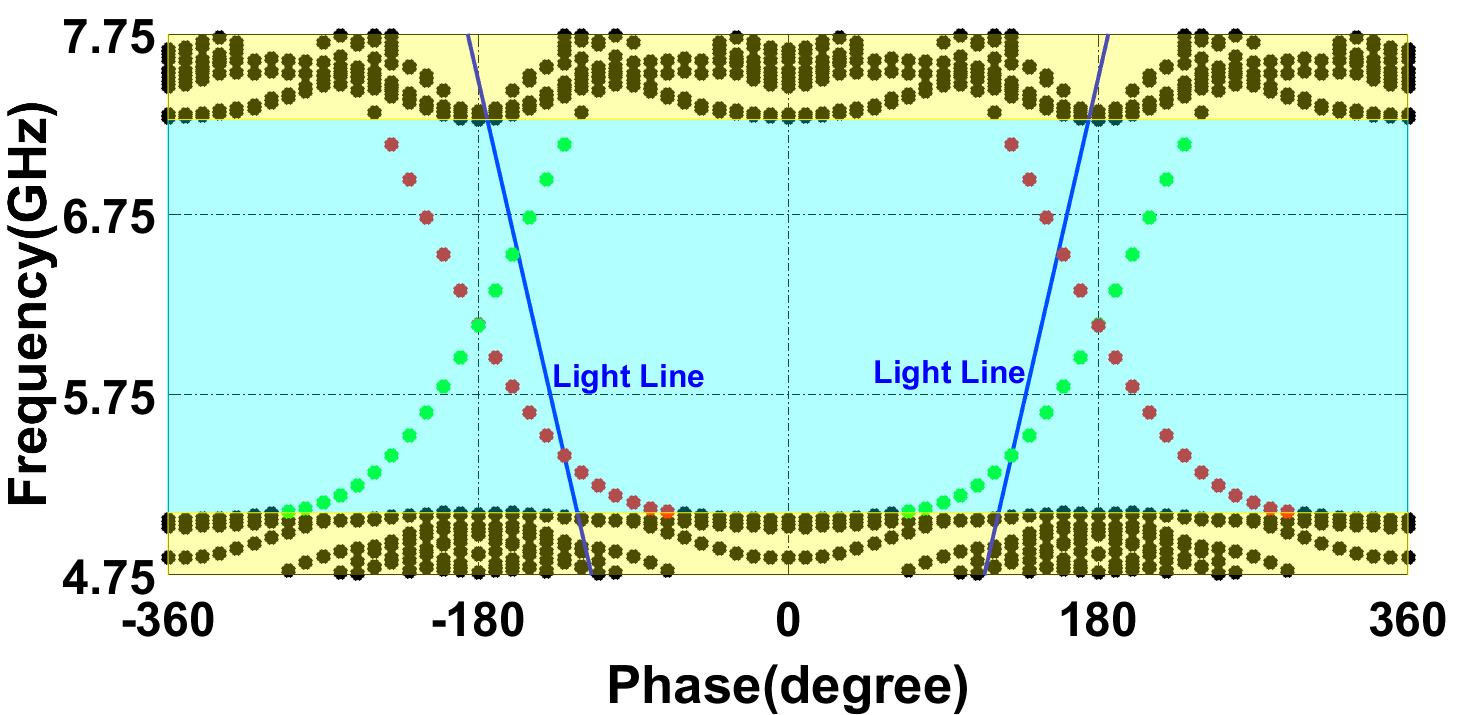}
	\label{Ribbon_Kagome_a_20_b_0433_h_157_SuperExtended2}
	}
	\end{center}
\caption{\subref{ZigZag_Ribbon_dispersion_a_20_b_0433_SuperExtended}-\subref{Ribbon_Kagome_a_20_b_0433_h_157_SuperExtended2} show the dispersion diagrams for a ribbon designed with three different configurations: a hexagonal unit cell arranged in a zigzag pattern, a 60-degree rhombic unit cell, and a kagome unit cell. The yellow region, marked with black dots, represents the bulk modes, while the cyan region, featuring red and green dots, indicates the edge modes. The differentiation of edge modes into red and green is attributed to the distinction between the two pseudospins. Additionally, the magenta region signifies the gap present in the edge modes. The 60-degree rhombic unit cell features an edge mode that has a bandgap, which is entirely located within the slow wave region. In contrast, the other configurations do not exhibit a bandgap and are suitable for leaky-wave antenna applications. 
  }
\label{Ribbon_dispersion}
\end{figure}

The initial indication of the topological nature of the kagome unit cell comes from the double degeneracy of the bands, similar to that described in \cite{bisharat2019}. To provide further evidence, the dispersion diagrams for one ribbon from each cell, as shown in Fig. \ref{Cell}, have been calculated. The results are presented in Figs. \ref{ZigZag_Ribbon_dispersion_a_20_b_0433_SuperExtended} to \ref{Ribbon_Kagome_a_20_b_0433_h_157_SuperExtended2}.
These results demonstrate the presence of a Dirac cone for the edge modes of both the zigzag and kagome ribbons, which is absent in the 60-degree rhombic ribbon.

Furthermore, despite the similarity between the 60-degree rombic and kagome cell in figure, the topological bandgap of the kagome cell is one third more than the 60 degree rhombic cell, approximately, which shows essential differences between them.

Unlike the 60-degree rhombic structure, the kagome structure exhibits no gap in the edge modes, So, a noteworthy advantage of the kagome lattice is its ability to simultaneously facilitate the continuous edge modes characteristic of the honeycomb lattice while also enabling the straight propagation observed in the rhombic lattice. Thus, this metasurface may be a suitable candidate for both guiding wave and leaky-wave antenna applications.

\begin{figure}[h!]
	\begin{center}
	\subfigure[]{\includegraphics[width=0.22\textwidth]{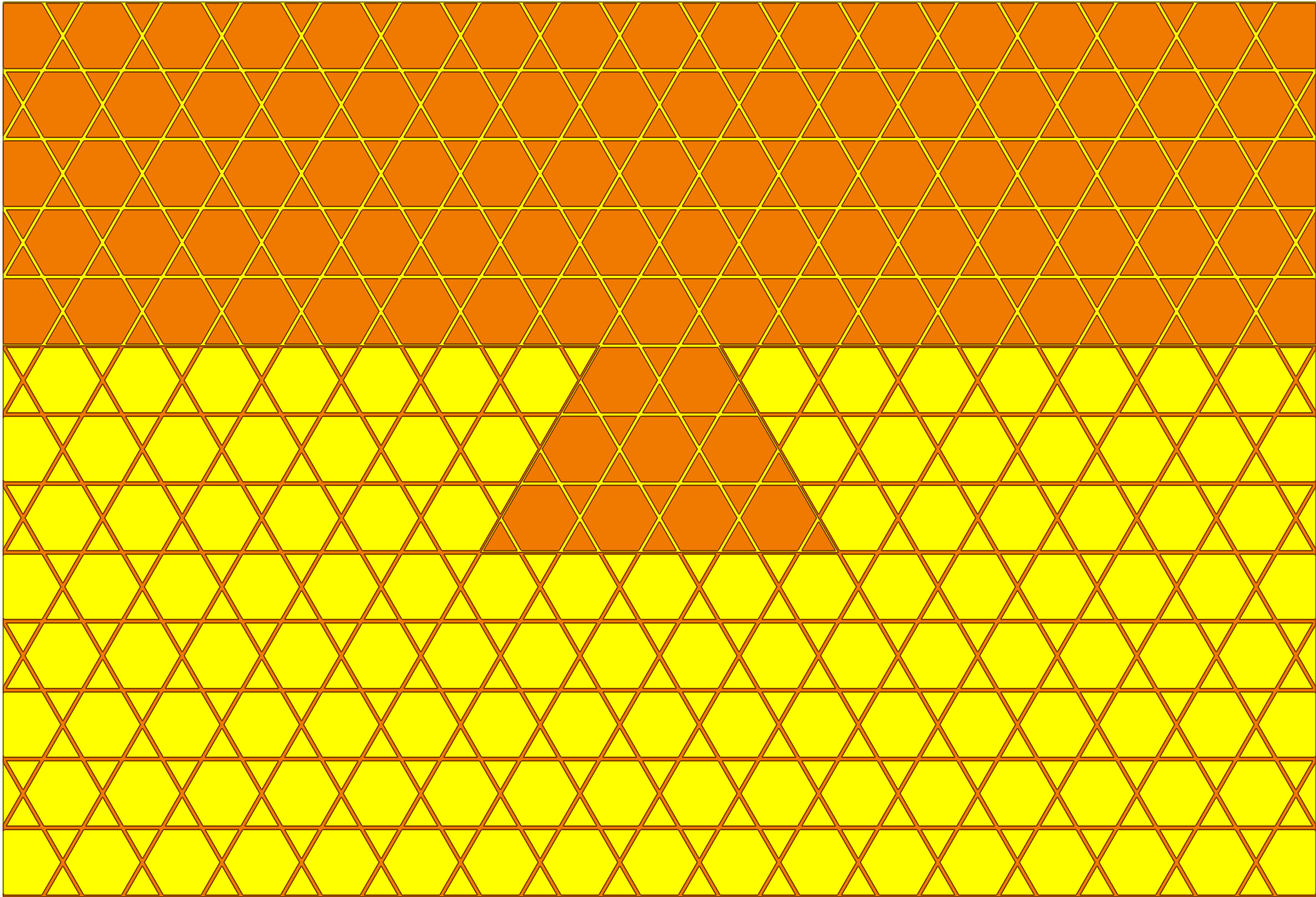}
	\label{Kagome_lattice2}
	}
	\hspace{0.00001cm}
	\subfigure[]{
	\includegraphics[width=0.22\textwidth]{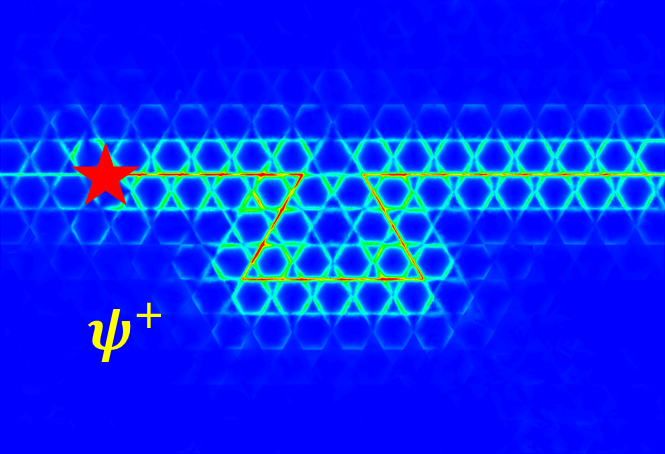}
	\label{prop}
	}
	\hfill
	\subfigure[]{
	\includegraphics[width=0.4\textwidth]{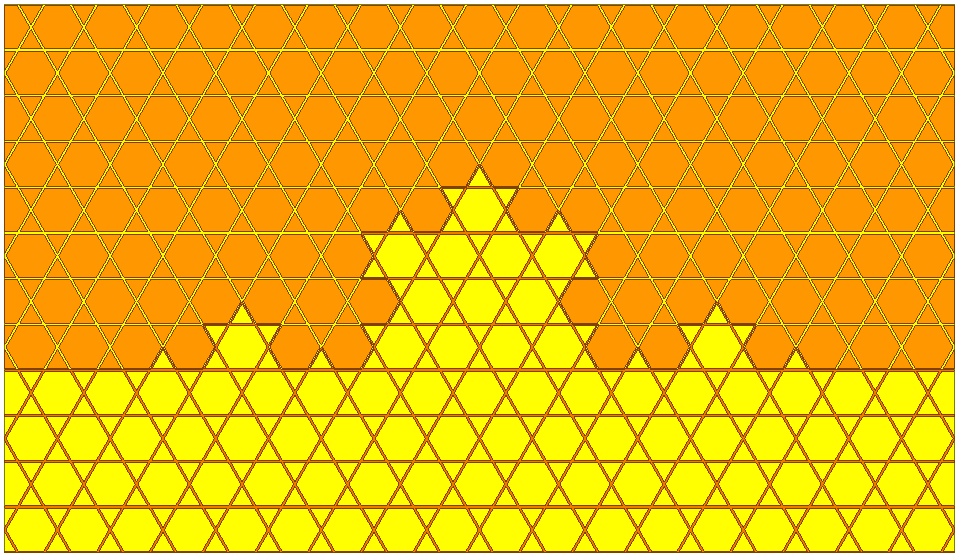}
	\label{Koch_snowflake}
	}
	\hspace{0.00001cm}
	\subfigure[]{
	\raisebox{0.15\height}{\includegraphics[width=0.22\textwidth]{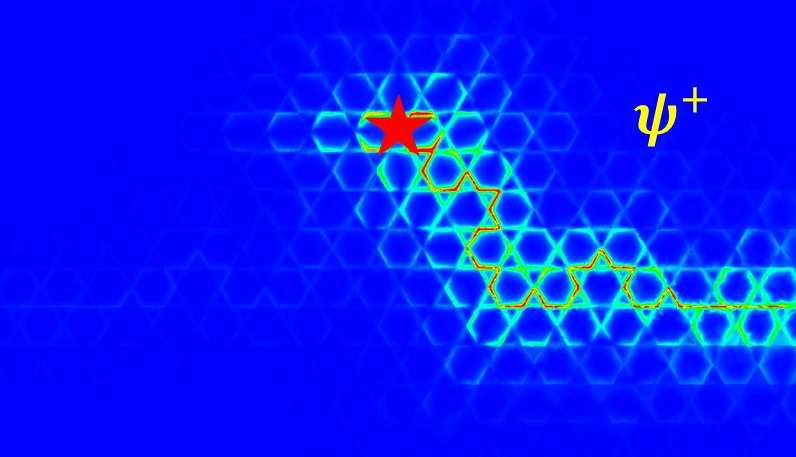}}
	\label{Koch_snowflake_Up}
	}
	\hspace{0.00001cm}
	\subfigure[]{
	\raisebox{0.15\height}{\includegraphics[width=0.22\textwidth]{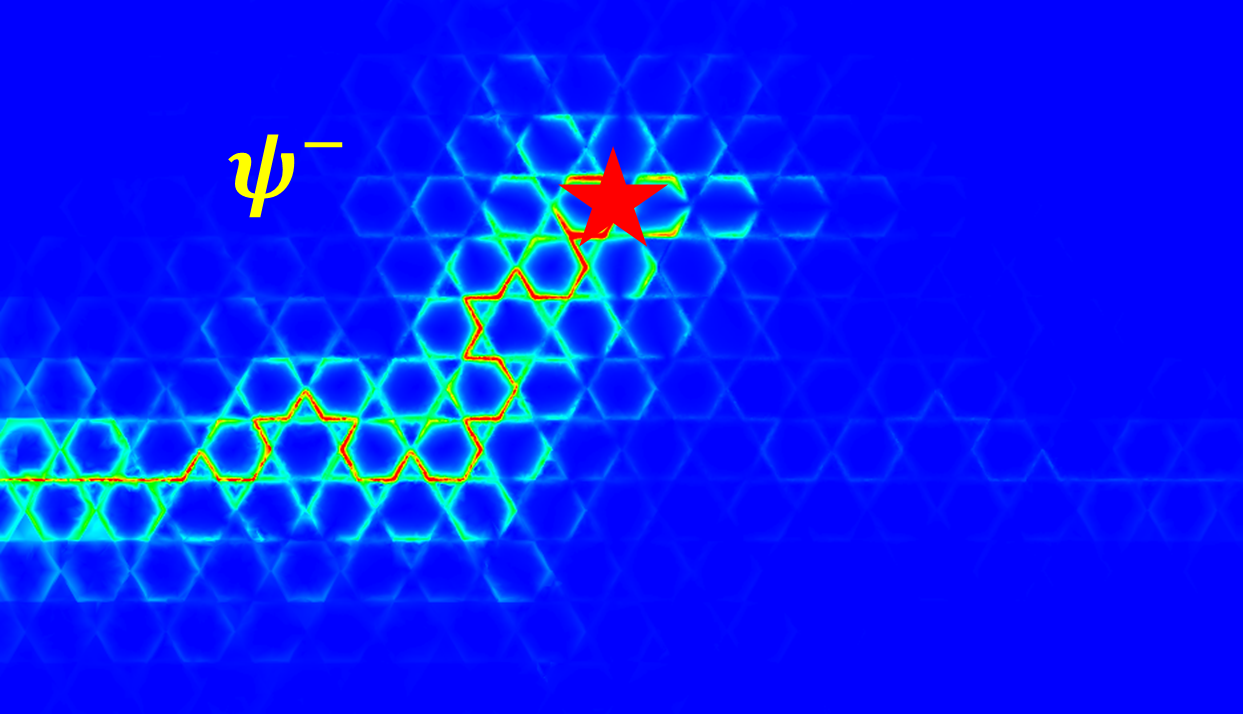}}
	\label{Koch_snowflake_Down}
	}
	\end{center}
\caption{
\subref{Kagome_lattice2} and \subref{prop} The structure is arranged in a kagome lattice, demonstrating the possibility for robust linear propagation.
\subref{Koch_snowflake} The shape of the interface has been changed to a Koch snowflake fractal, highlighting its ability to make sharp turns and its robustness in propagation. \subref{Koch_snowflake_Up} and \subref{Koch_snowflake_Down} demonstrate the unidirectional propagation of pseudospin up ( $\psi^+$ ) and down ( $\psi^-$ ), respectively. The excitation point is located at the snowflake's apex, marked by a red star.}
\label{propTotal}
\end{figure}

To validate the previous results, it is essential to observe the propagation of the edge mode along the interface line by exciting the pseudospins. The pseudospins are excited by a Hertzian dipole, represented by \(E_z + Z_0H_z\) and \(E_z - Z_0H_z\), where \(Z_0\) is the impedance of free space \cite{bisharat2019}. 

In the initial step to demonstrate the possibility of robust linear propagation, a pseudospin was excited in the structure depicted in Fig. \ref{Kagome_lattice2}, resulting in the robust propagation shown in Fig. \ref{prop}. This represents a significant advantage of the kagome lattice over the hexagonal lattice, as it allows for straight-line propagation, which is not feasible in the hexagonal structure.

To showcase the high degree of freedom and design flexibility of the structure, the shape of the interface line is designed as a Koch snowflake fractal (see Fig. \ref{Koch_snowflake}). The unidirectional propagation of these pseudospins is illustrated in Figs. \ref{Koch_snowflake_Up} and \ref{Koch_snowflake_Down} at a frequency of 5.4 GHz. These observations confirm previous findings and underscore the topological behavior of the kagome structure.

\subsection{Parametric study}
\begin{figure*}[h!]
	\begin{center}
	\subfigure[]{
	\includegraphics[width=0.4\textwidth]{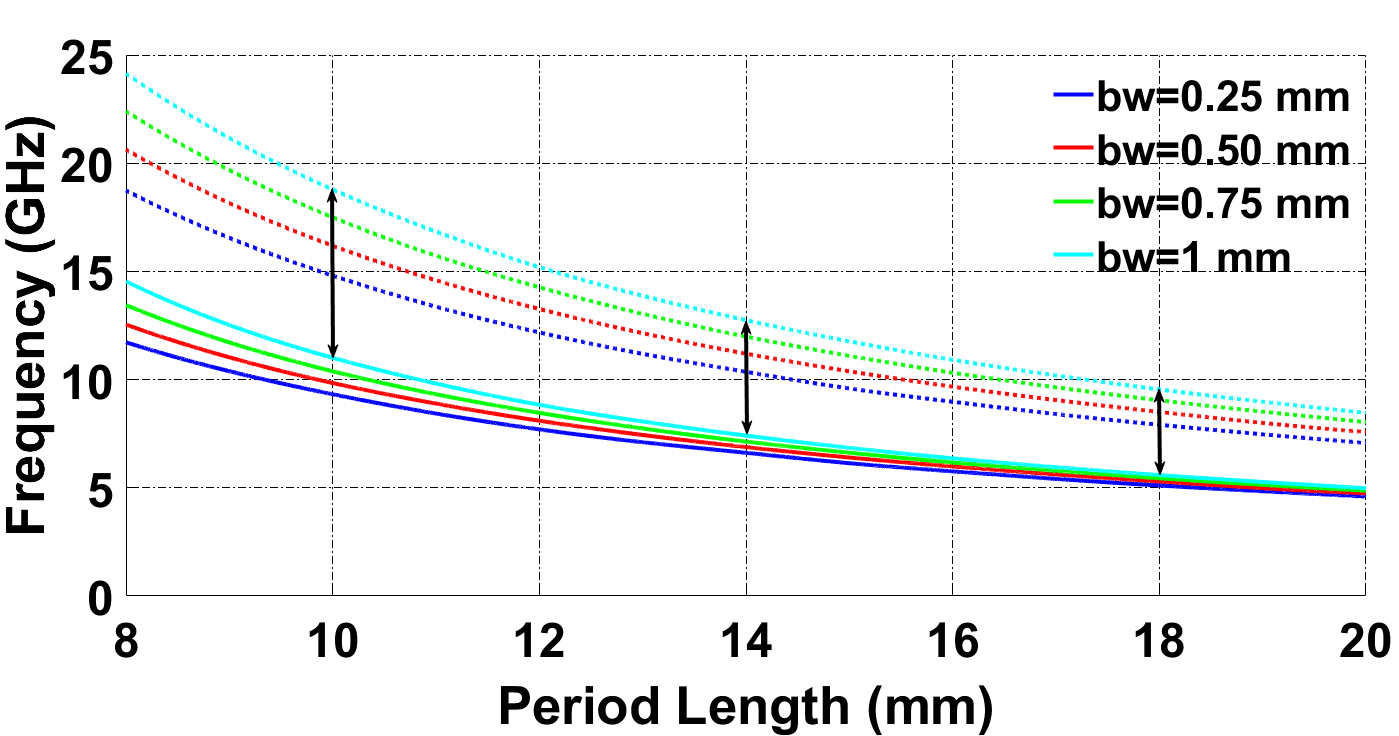}
	\label{two bands against a}
	}
	\hspace{0.001cm}
	\subfigure[]{
	\includegraphics[width=0.4\textwidth]{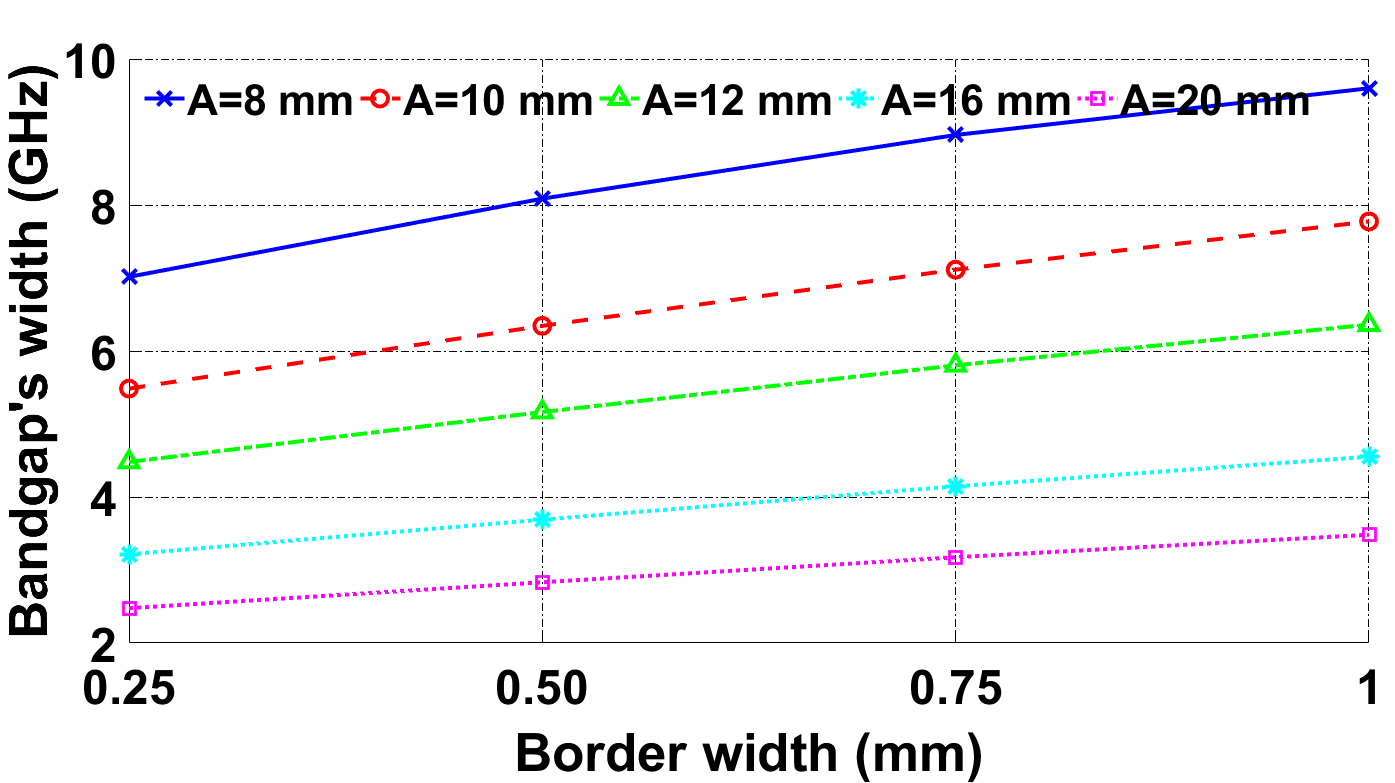}
	\label{bandgap aginst b}
	}
	\subfigure[]{
	\raisebox{0.1\height}{\includegraphics[width=0.3\textwidth]{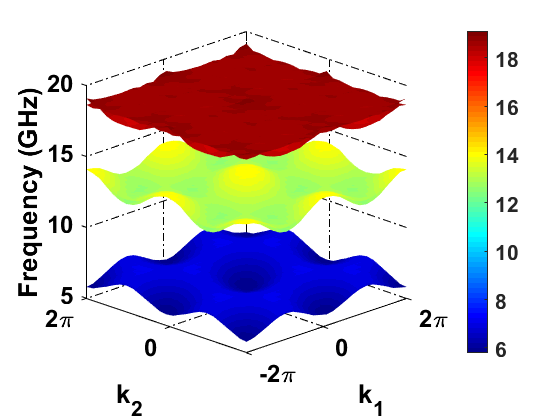}}
	\label{Kagome_3D_dispersion_13_0625}
	}
	\hspace{0.001cm}
	\subfigure[]{
	\includegraphics[width=0.5\textwidth]{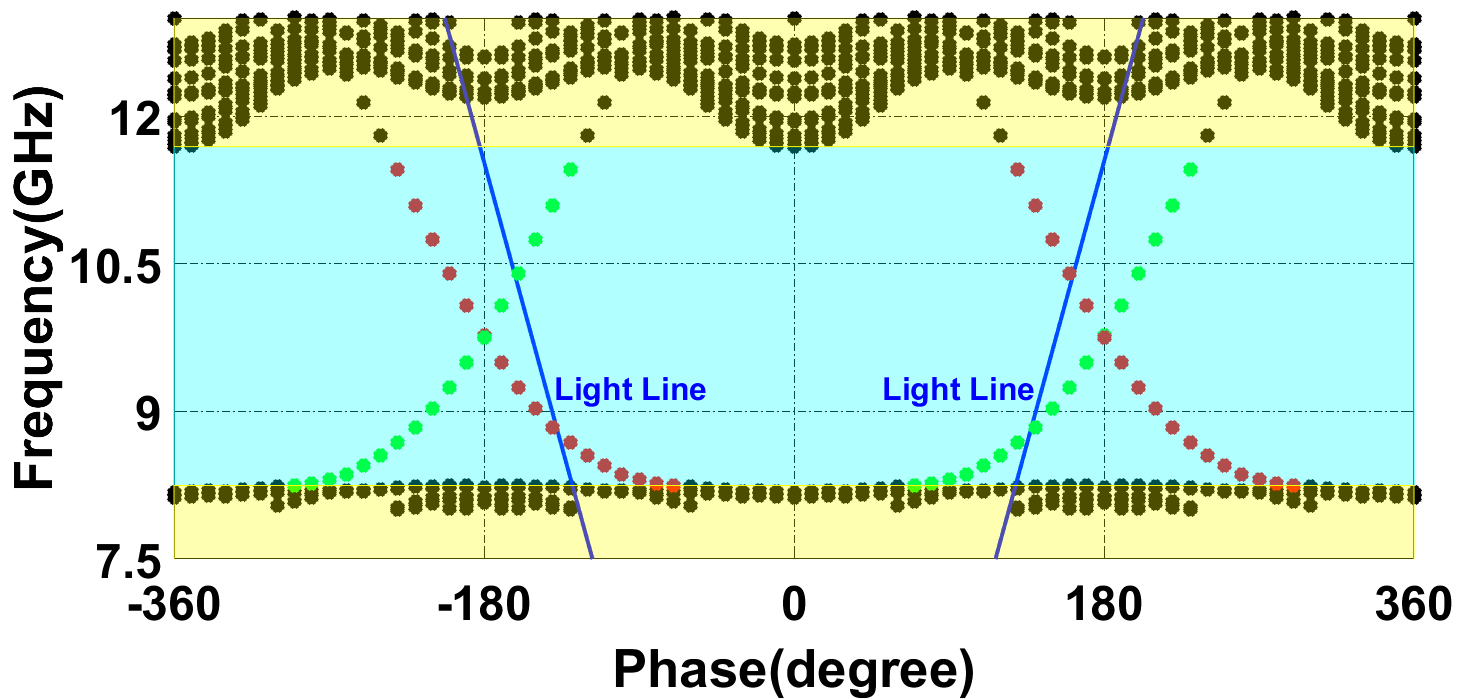}
	\label{Ribbon_Kagome_a_13_b_0625_h_157_SuperExtended2}
	}
	\end{center}
\caption{The parametric study of the kagome unit cell regarding period length and border width. \subref{two bands against a} The variation of the maximum of the lower band (solid lines) and the minimum of the upper band (dashed lines) with respect to the period length is demonstrated for various border widths. The arrows indicate the reduction in the bandgap width as the period length increases. \subref{bandgap aginst b}The variation of bandgap width in relation to border width is illustrated for several periodic lengths. It demonstrates that the width of the bandgap increases with an increase in border width or a decrease in periodic length. However, the impact of increasing the border width diminishes with an increase in the period length.
\subref{Kagome_3D_dispersion_13_0625} 
The 3D dispersion diagram of the kagome cell features a 13 mm period length and a border width of 0.625 mm. The substrate used is a 1.57 mm thick Rogers/Duroid 5880.
\subref{Ribbon_Kagome_a_13_b_0625_h_157_SuperExtended2} 
Dispersion diagram for a single ribbon of the kagome structure, based on the specified dimensions, reveals that a portion of the edge mode is situated in the fast wave region, which can be utilized for leaky-wave antenna applications. }
\label{parametric study}
\end{figure*}

 A parametric study is necessary to examine how the dimensions of the cell affect the limits of the topological bandgap. To achieve this, we calculated the dispersion diagram for various values of period length and border width. The results of this parametric study, illustrated in Fig. \ref{parametric study}, show that as the period length increases, the bandgap shifts to lower frequencies, and its width decreases. Conversely, increasing the border width results in a widening of the bandgap; however, the rate of increase depends on the period length, as it tends to decrease with longer period lengths.

Based on the results of the parametric study, the period length and border width were selected as 13 mm and 0.625 mm, respectively, to establish the topological bandgap in the X-band frequencies. To confirm that these dimensions achieve the desired frequency band, the dispersion diagram of one cell was calculated and is shown in Fig. \ref{Kagome_3D_dispersion_13_0625} in a 3D display, which validates the chosen dimensions. Additionally, to analyze the behavior of the edge mode, the dispersion diagram of one ribbon of the structure was obtained and is illustrated in Fig. \ref{Ribbon_Kagome_a_13_b_0625_h_157_SuperExtended2}. This dispersion diagram resembles that of Fig. \ref{Ribbon_Kagome_a_20_b_0433_h_157_SuperExtended2}, with some portions of the mode located in the fast wave region, making it suitable for leaky-wave antenna applications.

\section{Leaky-wave antenna application}


	As demonstrated in previous sections, the kagome structure provides substantial flexibility for incorporating arbitrary turns and ensures robust propagation. These features enable the modification of the interface line configuration to resemble the armchair layout described in \cite{abtahiArmchair}, where the armchair arrangement of hexagonal unit cells is utilized in the configuration of the interface line.

\begin{figure}[b!]
	\begin{center}

	\raisebox{0.22\height}{\includegraphics[width=0.08\textwidth]{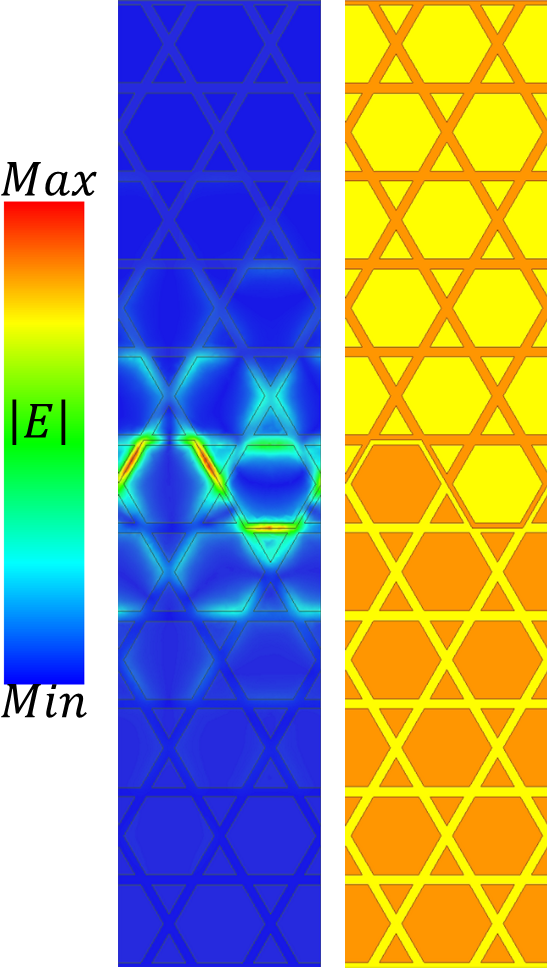}}
	\includegraphics[width=0.35\textwidth]{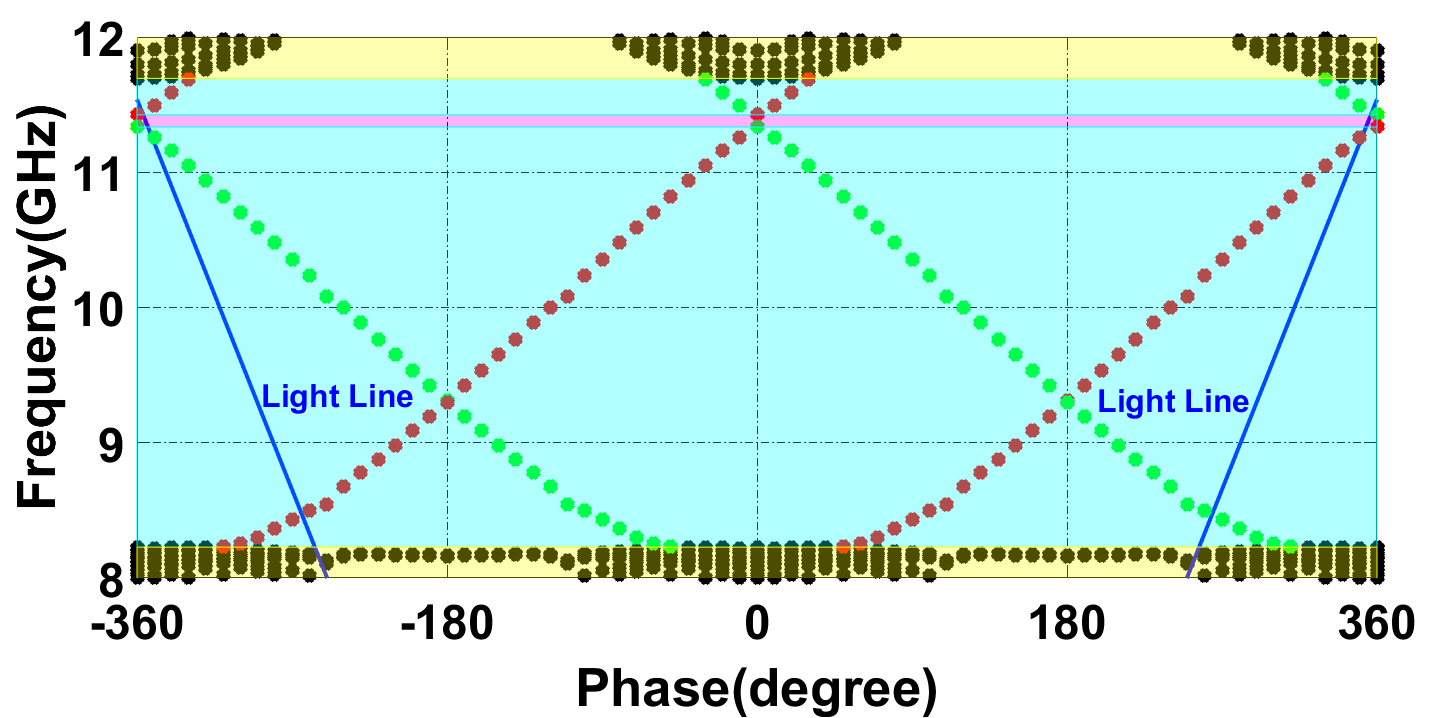}

	\end{center}
\caption{ 
The dispersion diagram of the demonstrated kagome structure ribbon in the armchair arrangement shows the presence of two modes in the fast wave region simultaneously. One of these modes exhibits right-hand behavior due to its positive phase velocity, while the other mode displays left-hand behavior, corresponding to its negative phase velocity.
}
\label{Kagome_Hex_Ribbon}
\end{figure}

Using the armchair arrangement has two main effects. First, it reduces the slope of the light line and expands the fast wave region for lower frequencies. Second, it affects the behavior of the edge mode and its spatial harmonics. A ribbon of the design is illustrated in Fig. \ref{Kagome_Hex_Ribbon}. The presence of the electric field at the interface confirms the accuracy of our calculations and the identification of the edge modes.

The dispersion diagram for the ribbon, shown in Fig.\ref{Kagome_Hex_Ribbon}, reveals the presence of an edge mode similar to that of the hexagonal configuration described in \cite{abtahiArmchair}. This edge mode spans the entire bandgap, indicating that this structure can be utilized for a leaky-wave antenna operating throughout the topological bandgap. This capability will influence the antenna's bandwidth and scanning range.

A critical aspect of the dispersion diagram is the positioning of the edge modes and their spatial harmonics in the fast wave region. This positioning indicates the simultaneous propagation of two modes with positive and negative phase velocities. In contrast to the findings presented in \cite{abtahiArmchair}, which consider the placement of these spatial harmonics in the fast wave region to be problematic, the proposed kagome structure views it as a beneficial feature. This configuration enables radiation to occur in both forward and backward directions, allowing the leaky-wave antenna to generate two beams directed forward and two beams directed backward.

\begin{figure}[t!]
	\begin{center}
	\subfigure[]{
	\includegraphics[width=0.45\textwidth]{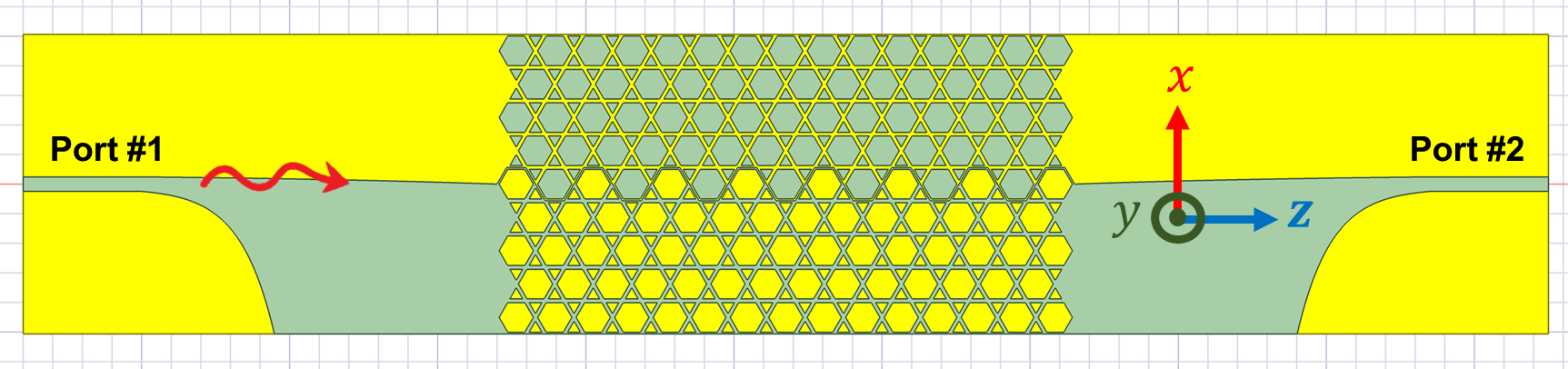}
	\label{kagome_Hex_Display_coor}
	}
	\hfill
	\subfigure[]{
	\includegraphics[width=0.45\textwidth]{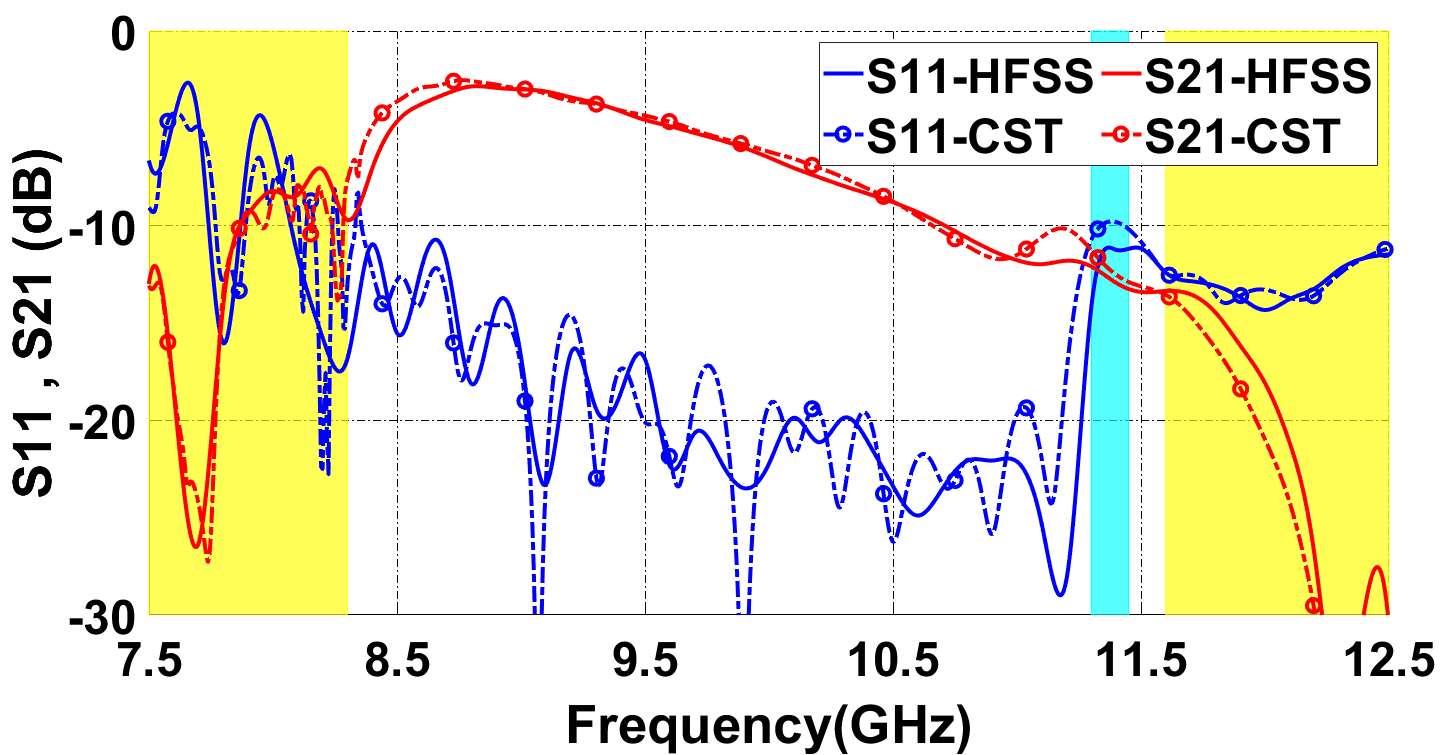}
	\label{Kagome_hex_S_parameters_13}
	}
	\end{center}
\caption{ 
The structure of the proposed leaky-wave antenna using a kagome lattice arranged in an armchair configuration is shown in \subref{kagome_Hex_Display_coor}. The antenna is excited using an ASL, similar to the approach in \cite{abtahiArmchair}. In \subref{Kagome_hex_S_parameters_13}, the return and insertion losses from simulations conducted in Ansys HFSS and CST Studio are presented and aligned, confirming the accuracy of the simulation results. The yellow bands indicate the presence of bulk modes, while the cyan band identifies the gap region in Fig.\ref{Kagome_Hex_Ribbon}, where the return loss has increased.
}
\label{S_para}
\end{figure}

The antenna structure is illustrated in Fig. \ref{kagome_Hex_Display_coor} and is connected to the classical line via the Antipodal Slot Line (ASL), as detailed in \cite{davisASL_AIP}. Due to the armchair shape of the interface, the connection between the ASL and the topological structure is similar to the discussion presented in \cite{abtahiArmchair}. This connection will help address the cavity-like effect caused by fragmented cells in the transition region between the ASL and the topological structure.

The antenna structure consists of four periods of patches or grids on each side. While using more periods could gradually create an ideal periodic structure, practical implementation would be challenging, if not impossible. However, as demonstrated in \cite{SinghAntenna}, three periods are sufficient to meet Bloch's condition for periodic cells. Therefore, choosing four periods is appropriate, and increasing the number of periods on either side will not significantly affect the field enhancement at the interface. As shown in Fig.\ref{Ribbon_Kagome_a_20_b_0433_h_157_SuperExtended2} and \ref{Kagome_Hex_Ribbon}, the weak confinement of the field at the interface is an inherent characteristic of the proposed kagome structure, making it an optimal candidate for leaky-wave antenna applications.

The return and insertion losses of the proposed antenna, simulated using Ansys HFSS and CST Studio, are presented in Fig. \ref{Kagome_hex_S_parameters_13}. The results from both simulations are in good agreement, validating the findings. The frequency ranges for the bulk modes and the edge mode bandgap are highlighted in yellow and cyan, respectively. These results align with the dispersion diagram in Fig.\ref{Kagome_Hex_Ribbon}, validating the simulations.

\begin{figure}[t!]
	\begin{center}
	\subfigure[]{
	\includegraphics[width=0.47\textwidth]{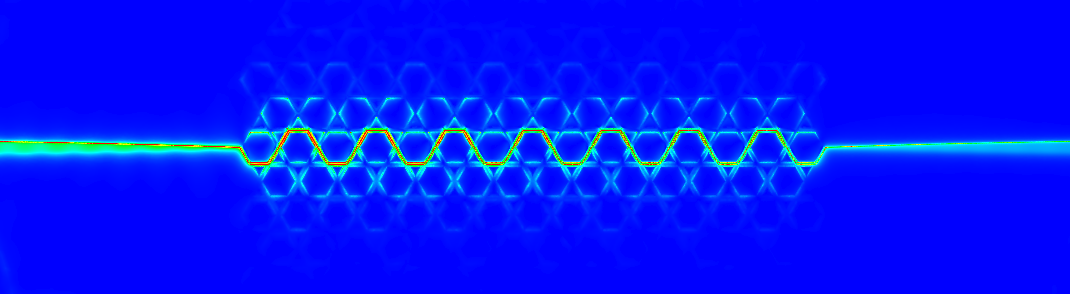}
	\label{Kagome_Hex_10GHz}
	}
	\hfill
	\subfigure[]{
	\includegraphics[width=0.47\textwidth]{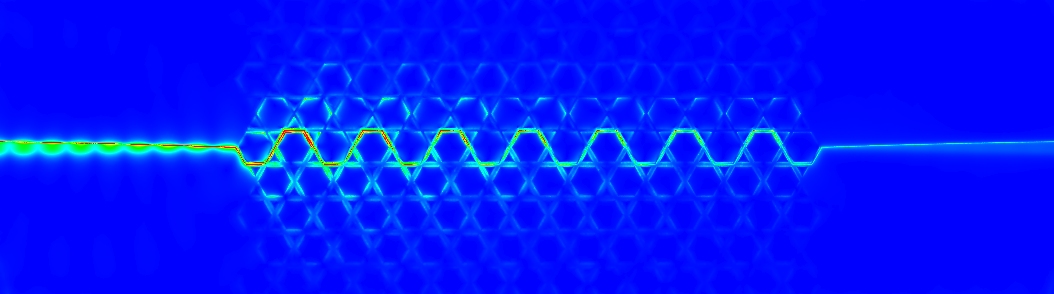}
	\label{Kagome_Hex_11.5GHz}
	}
	\end{center}
\caption{ 
The electrical field profile of the structure at frequencies of 10 GHz (\subref{Kagome_Hex_10GHz}) and 11.5 GHz (\subref{Kagome_Hex_11.5GHz}) reveals that the attenuation of the bulk modes at 11.5 GHz is significantly lower than at 10 GHz. As a result, the radiation pattern at 11.5 GHz is not acceptable. These findings are consistent with the dispersion diagram shown in Fig.\ref{Kagome_Hex_Ribbon} and the S-parameters illustrated in Fig.\ref{Kagome_hex_S_parameters_13}, which validate the simulations.
}
\label{Evanescent}
\end{figure}

The frequency range of 11.1 to 11.7 GHz cannot be selected, even though it falls within the topological bandgap. This issue arises due to a reduction in the attenuation of bulk modes within that range. To illustrate this problem visually, the electric field profiles within the structure are shown in Fig. \ref{Evanescent} for frequencies of 10 GHz and 11.5 GHz. At 10 GHz, the electric field is confined to the interface line and decreases rapidly through the bulk; however, at 11.5 GHz, the attenuation rate of the bulk modes decreases significantly and cannot be overlooked. These bulk modes would negatively impact the radiation pattern, making it undesirable. Therefore, this frequency range is not operational and the desired operating frequency range of 8.8-11.1 GHz should been selected to facilitate forward and backward radiation. 

\begin{figure*}[t!]
	\begin{center}
	\subfigure[]{
	\includegraphics[width=0.7\textwidth]{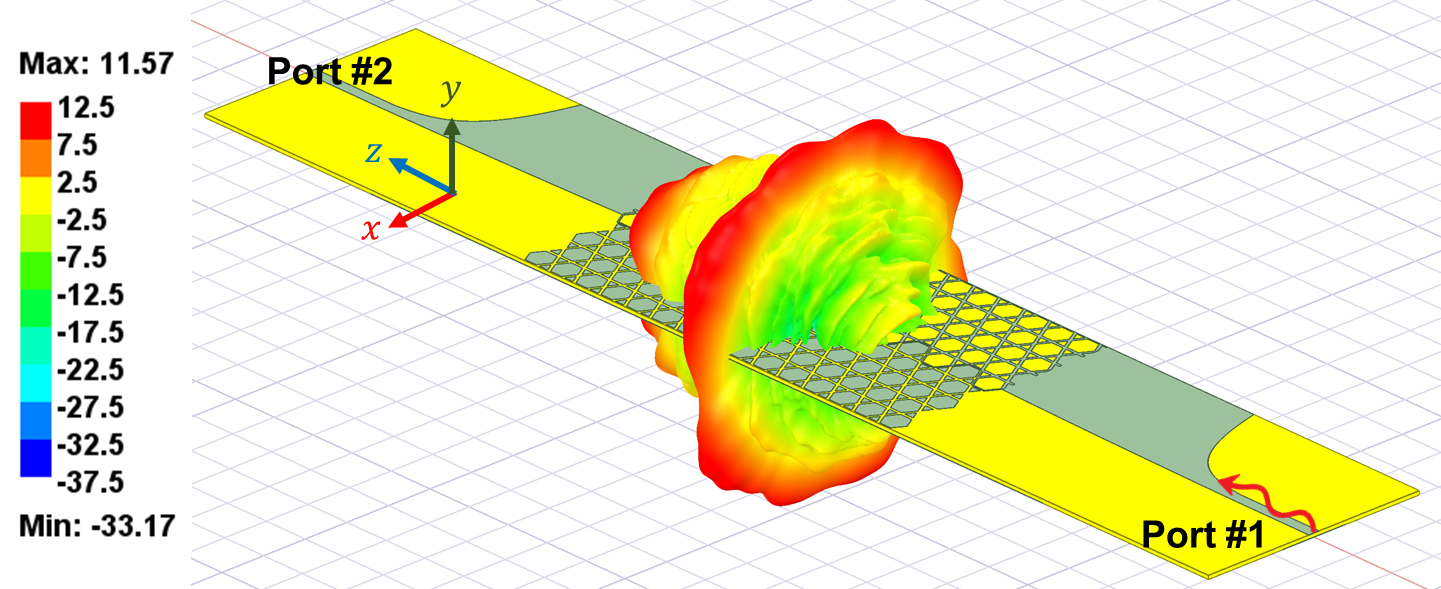}
	\label{3D_pattern_coor}
	}
	\hspace{0.0001cm}
	\subfigure[]{
	\raisebox{0.0\height}{\includegraphics[width=0.42\textwidth]{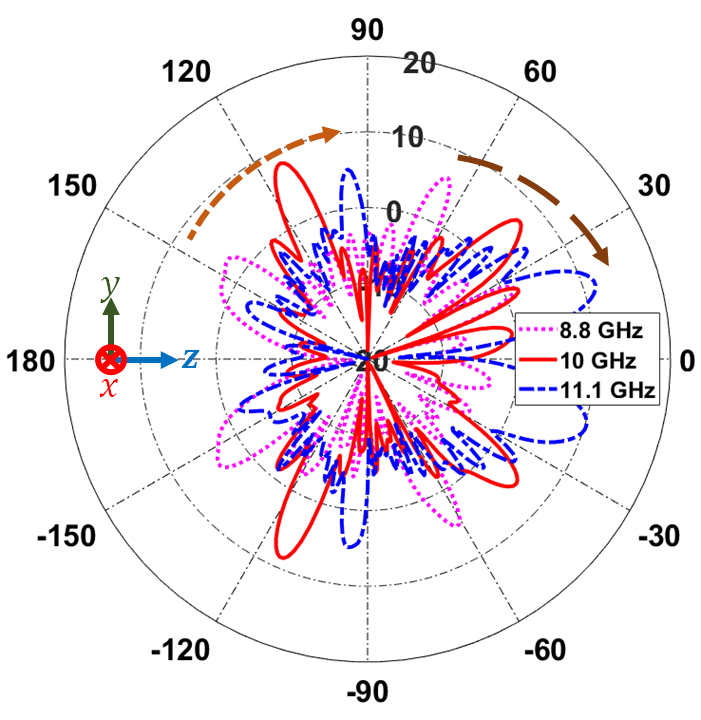}}
	\label{Plar_realized_gain_HFSS_coor}
	}	
	\hspace{0.0001cm}
	\subfigure[]{
	\raisebox{0.0\height}{\includegraphics[width=0.42\textwidth]{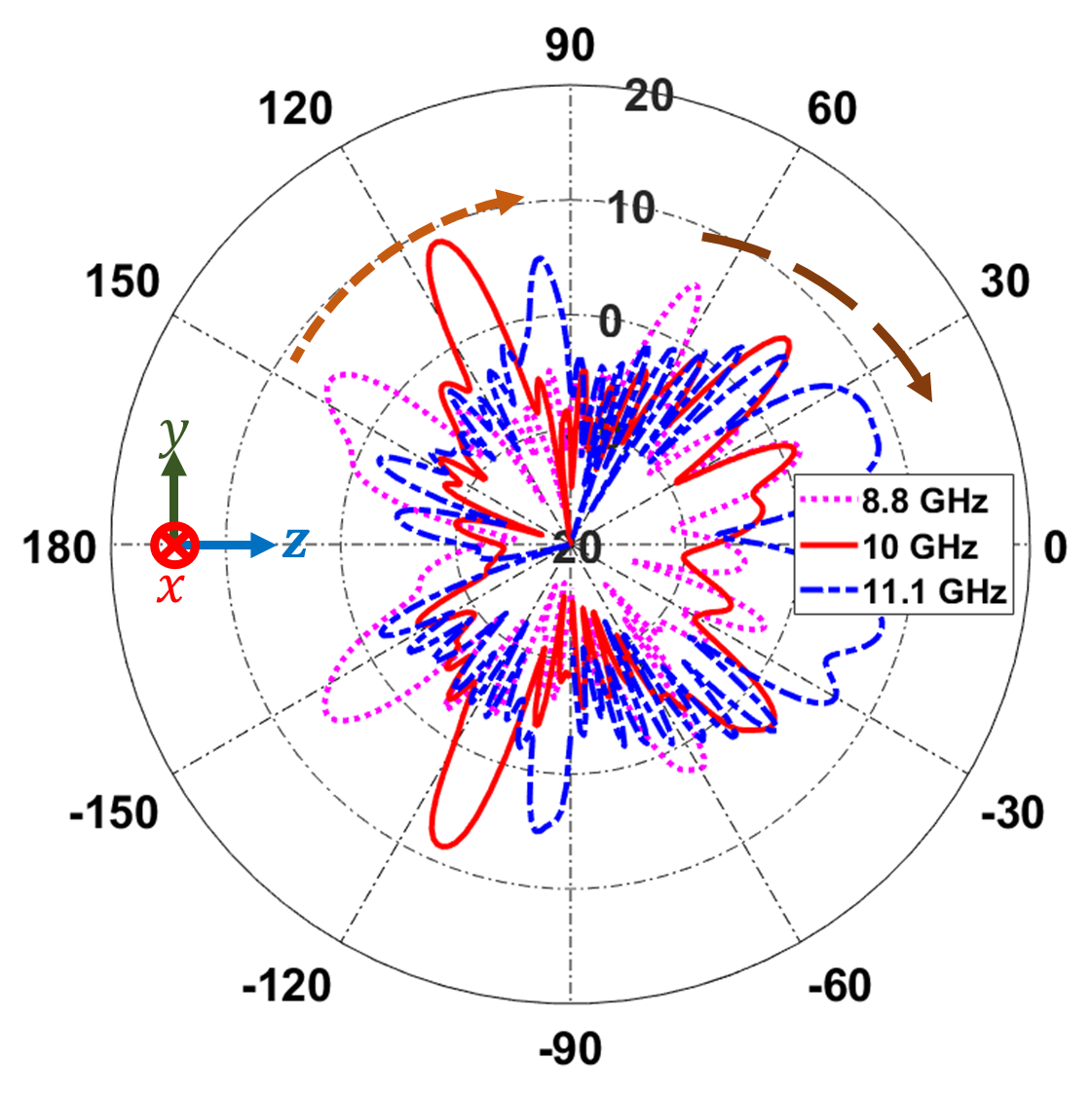}}
	\label{Plar_realized_gain_CST_coor}
	}		
	\hfill
	\subfigure[]{
	\raisebox{0.0\height}{\includegraphics[width=0.31\textwidth]{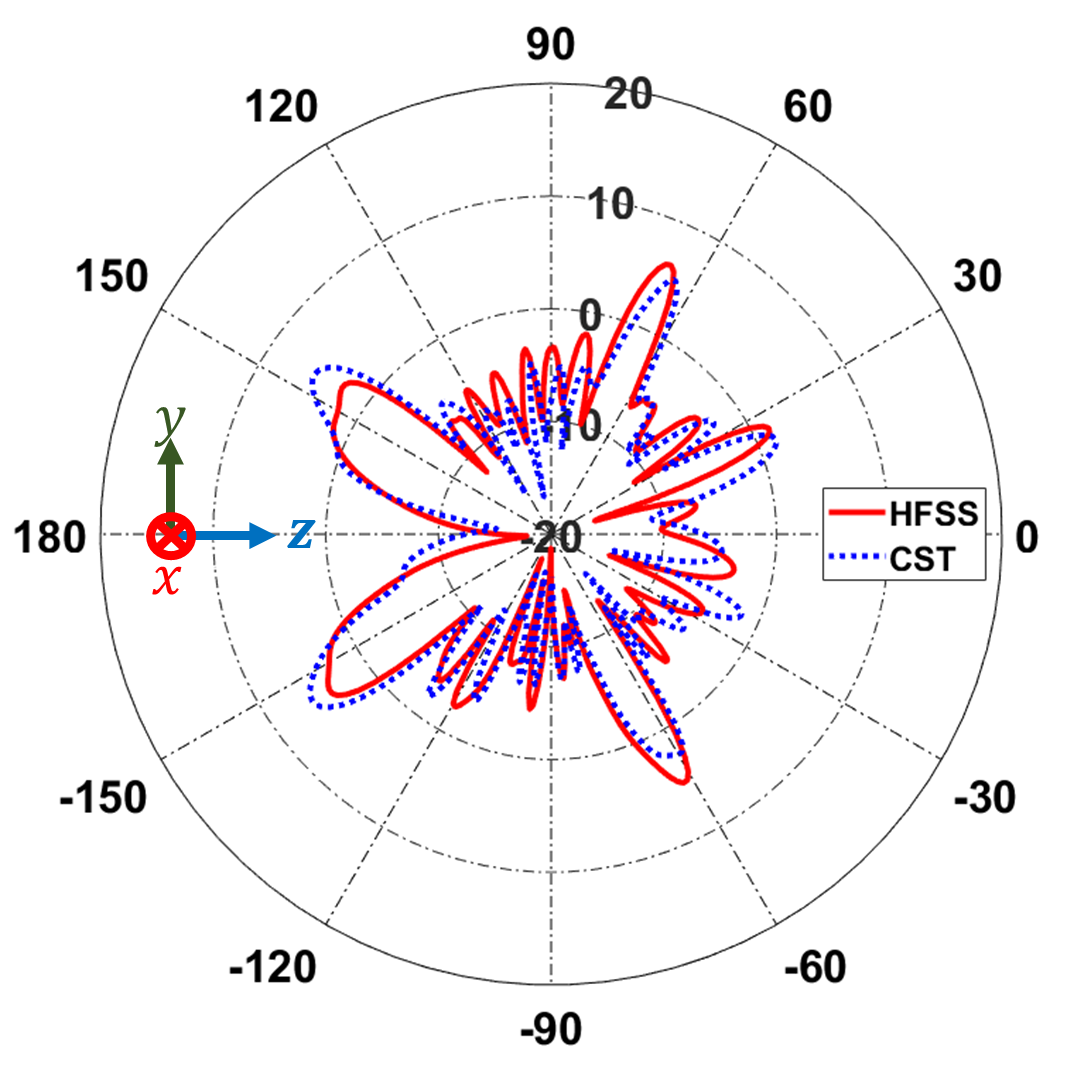}}
	\label{Plar_realized_gain_8.8_coor}
	}			
	\hspace{0.0001cm}
	\subfigure[]{
	\raisebox{0.0\height}{\includegraphics[width=0.31\textwidth]{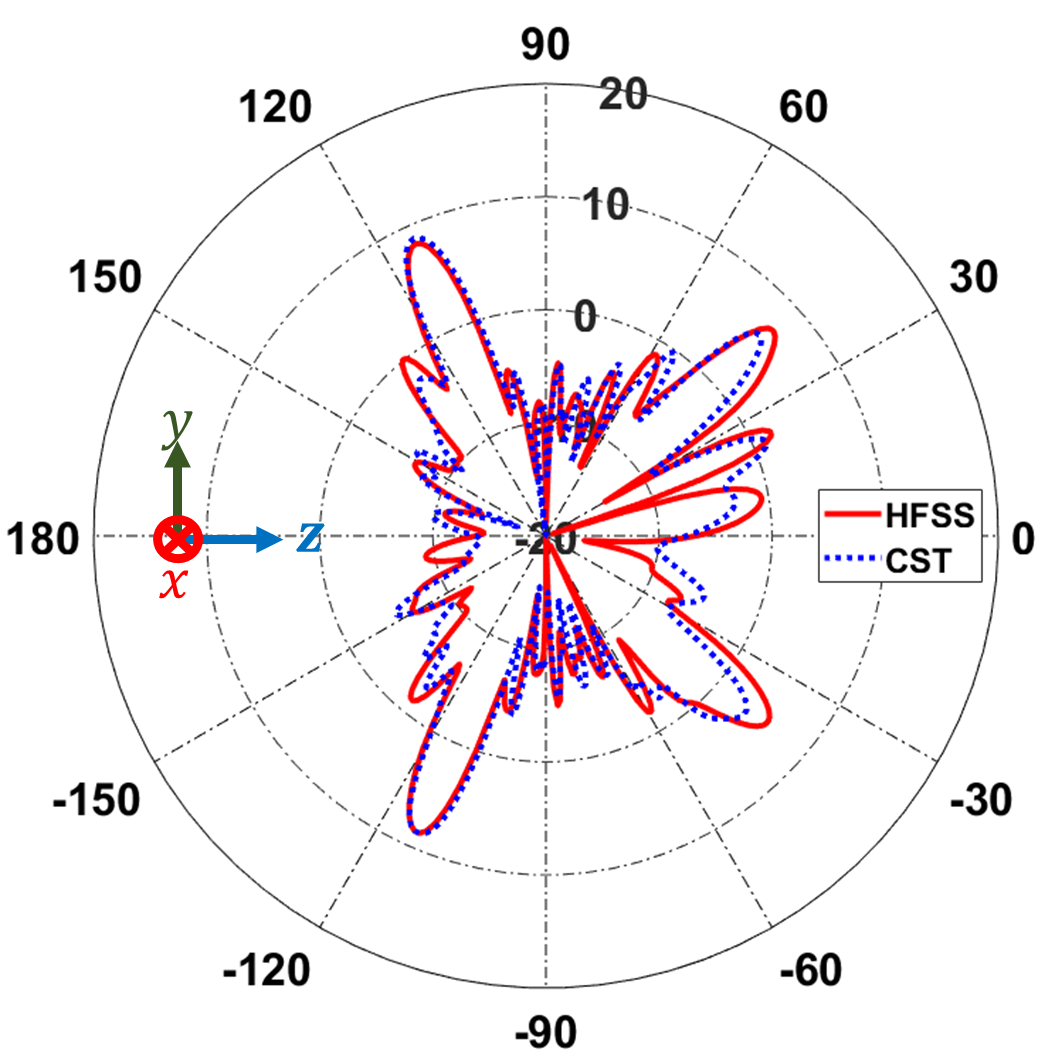}}
	\label{Plar_realized_gain_10_coor}
	}			
	\hspace{0.0001cm}
	\subfigure[]{
	\raisebox{0.0\height}{\includegraphics[width=0.31\textwidth]{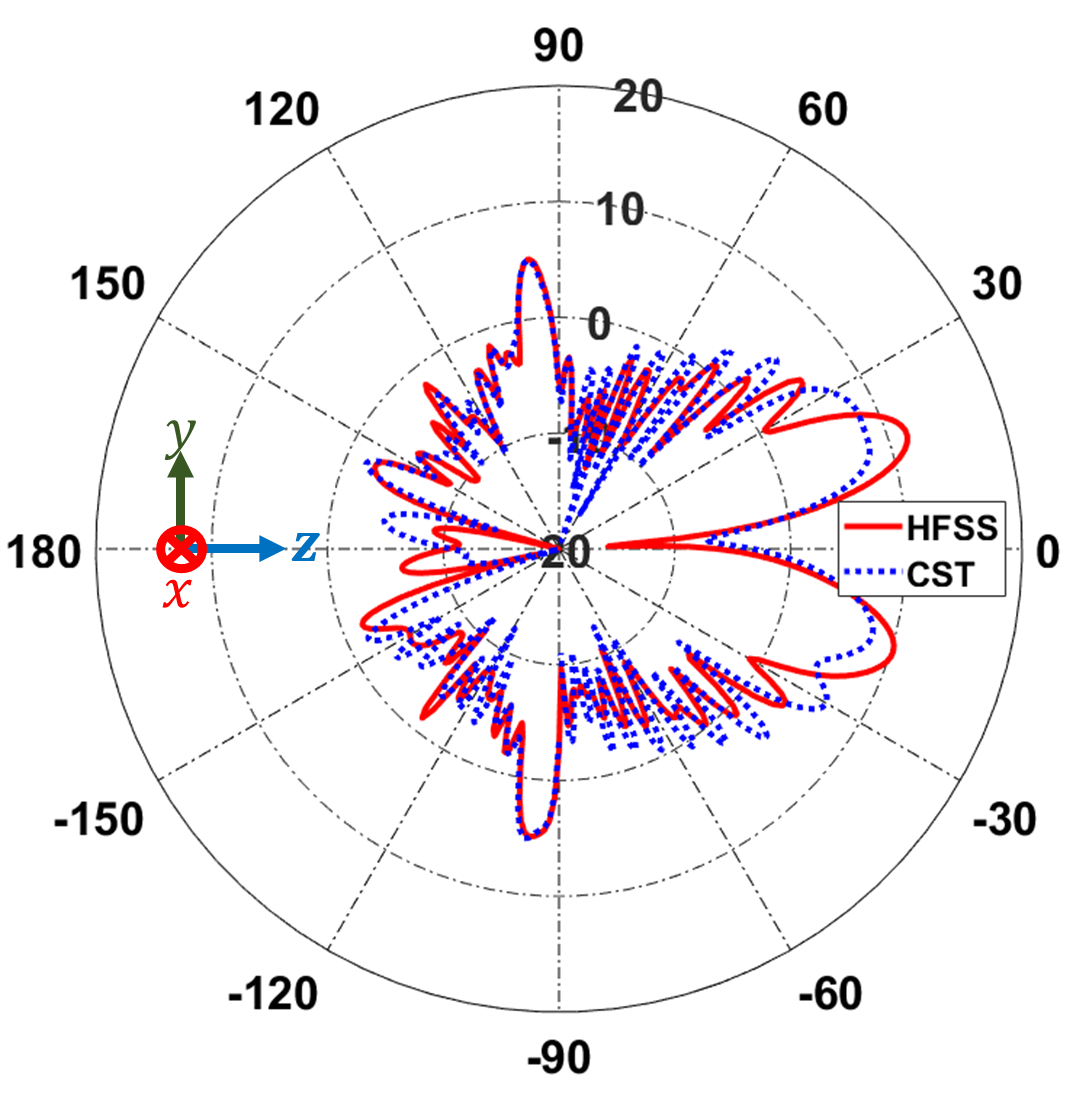}}
	\label{Plar_realized_gain_11.1_coor}
	}
	\end{center}
\caption{\subref{3D_pattern_coor} 
The realized gain pattern of the proposed antenna at 11 GHz shows two beams directed forward and two directed backward. The pair of backward beams is positioned near the broadside, while the forward beams are located close to the endfire direction.
\subref{Plar_realized_gain_HFSS_coor} and \subref{Plar_realized_gain_CST_coor} 
represent the realized gain pattern results for \(\phi = 90^\circ\) in Ansys HFSS and CST Studio, respectively, at frequencies of 8.8 GHz, 10 GHz, and 11.1 GHz. The arrows indicate the direction of rotation for the main lobes as the frequency increases. 
The comparison of results from Ansys HFSS and CST Studio, as shown in \subref{Plar_realized_gain_8.8_coor}-\subref{Plar_realized_gain_11.1_coor} for frequencies of 8.8 GHz, 10 GHz, and 11.1 GHz, clearly establishes a strong alignment between the simulations from both software platforms. This robust agreement in the observed patterns decisively confirms the reliability of the results.
}
\label{Patterns}
\end{figure*}

Based on prior analysis, the radiation pattern of the proposed antenna was calculated and is illustrated in Fig. \ref{Patterns}. The 3D visualization of the realized gain pattern at 11 GHz, shown in Fig. \ref{3D_pattern_coor}, highlights the antenna’s ability to generate both forward and backward beams. However, Due to the reduced attenuation of bulk modes at higher frequencies within the topological bandgap (Fig.\ref{Kagome_Hex_11.5GHz}), the proposed antenna does not facilitate radiation in the broadside direction. This behavior is consistent with the dispersion diagram presented in Fig. \ref{Kagome_Hex_Ribbon}.

\renewcommand{\arraystretch}{1.6}
\begin{table*}[t!]
	\begin{center}
		\caption{comparison between leaky-wave antennas based on PTIs.}
		\begin{tabular}{cccccccc}
				 				& \multirow{ 2}{*}{Type} & Lattice     & \multirow{ 2}{*}{Bandwidth}       & Number of        & \multirow{ 2}{*}{HPBW$^1$}  & \multirow{ 2}{*}{Scanning Range$^2$}     & \multirow{ 2}{*}{Length}   \\
				 				&   & Interface shape     &         &  beams  &       &      &    \\ \hline 
			\multirow{ 2}{*}{\cite{chernAntenna}} & \multirow{ 2}{*}{Chern PTI} & square & \multirow{ 2}{*}{6.34-6.76 GHz}   & \multirow{ 2}{*}{1} & \multirow{ 2}{*}{$4.5^{\circ}$} & \multirow{ 2}{*}{$\approx 70^{\circ}$ to $110^{\circ}$}& \multirow{ 2}{*}{525 mm}    \\
			 &  & Straight &    & & & &     \\ \hline
			\multirow{ 2}{*}{\cite{ChernPLWA}} & \multirow{ 2}{*}{Chern PTI} & square & \multirow{ 2}{*}{13.2-14 GHz }  & \multirow{ 2}{*}{1} & \multirow{ 2}{*}{$\approx 10^{\circ}$} & \multirow{ 2}{*}{$\approx 52^{\circ}$ to $40^{\circ}$}& \multirow{ 2}{*}{156 mm}     \\
			 &  & Straight &    &  & & &     \\ \hline
			
			\multirow{ 2}{*}{\cite{valleyLWA} } & \multirow{ 2}{*}{Valley PTI} & Hexagonal  & \multirow{ 2}{*}{8.2-8.7 GHz} & \multirow{ 2}{*}{1} & \multirow{ 2}{*}{$\approx 7^{\circ}$}& \multirow{ 2}{*}{$\approx 105^{\circ}$ to $80^{\circ}$ } & \multirow{ 2}{*}{264 mm }  \\ 
								&		& Even-mode	&				  &				  &	 &   &            \\ \hline
			
			\multirow{ 2}{*}{\cite{SinghAntenna} } & \multirow{ 2}{*}{Spin PTI} & Hexagonal  & \multirow{ 2}{*}{19.75-21.25 GHz} & \multirow{ 2}{*}{2} & \multirow{ 2}{*}{$\approx 6^{\circ}$} & \multirow{ 2}{*}{$\pm166^{\circ}$ to $\pm130^{\circ}$}  & \multirow{ 2}{*}{$\approx$168 mm }  \\ 
								&		& Zig-Zag	&				  &		&		  &	   &            \\ \hline
								
			\multirow{ 2}{*}{\cite{abtahi2} }& \multirow{ 2}{*}{Spin PTI} & 30$^\circ$ Rhombic & \multirow{ 2}{*}{17.5-20 GHz} & \multirow{ 2}{*}{2} & \multirow{ 2}{*}{$ 5^{\circ}$} & \multirow{ 2}{*}{$\pm122^{\circ}$ to $\pm99.5^{\circ}$}& \multirow{ 2}{*}{200 mm}   \\ 
								&			& Straight & 	 &			&	  &     & 	    \\ \hline
								
			\multirow{ 2}{*}{\cite{abtahiArmchair} }& \multirow{ 2}{*}{Spin PTI} & Hexagonal & \multirow{ 2}{*}{8.1-10.8 GHz} & \multirow{ 2}{*}{2} & \multirow{ 2}{*}{$ 18^{\circ}$} & \multirow{ 2}{*}{$\pm126^{\circ}$ to $\pm73^{\circ}$}& \multirow{ 2}{*}{97 mm}   \\ 
								&			&  Armchair & 	  &		&		  &        & 	    \\ \hline
								
			\multirow{ 2}{*}{This work }& \multirow{ 2}{*}{Spin PTI} & Kagome & \multirow{ 2}{*}{8.8-11.1 GHz} & \multirow{ 2}{*}{4} & Backward : $6^{\circ}$ &  $\pm145^{\circ}$ to $+\pm95^{\circ}$& \multirow{ 2}{*}{$\approx$195 mm}   \\ 
								&			&  Armchair & 	  &			&	Forward : $11^{\circ}$  &  $\pm64^{\circ}$ to $\pm17^{\circ}$& 	    \\ \hline
			\multicolumn{8}{l}{$^1$ The value that corresponds to the maximum gain within the scan range is reported.}\\
			\multicolumn{8}{l}{$^2$ The broadside is at $90^{\circ}$}\\
			
		\end{tabular}
		\label{tableC}
	\end{center}
\end{table*}

The scanning behavior of the antenna was analyzed using Ansys HFSS and CST Studio, with results presented in Figures \ref{Plar_realized_gain_HFSS_coor} and \subref{Plar_realized_gain_CST_coor}, respectively. These figures illustrate the realized gain patterns in the plane of $\phi = 90^\circ$ for frequencies of 8.8 GHz, 10 GHz, and 11 GHz. The backward beams are observed to shift from approximately $\pm145^{\circ}$ to $\pm95^{\circ}$, while the forward beams shift from about $\pm64^{\circ}$ to $\pm17^{\circ}$. Consequently, the forward and backward beams exhibit scan ranges of $50^\circ$ and $47^\circ$, respectively. The proximity of these two scan ranges aligns with the edge modes shown in the dispersion diagram of Figure \ref{Kagome_Hex_Ribbon}, thereby confirming these results.\\
Moreover, to demonstrate the consistency between the simulation results from Ansys HFSS and CST Studio, the realized gain patterns for 8.8 GHz, 10 GHz, and 11.1 GHz are plotted in separate figures: Fig.\ref{Plar_realized_gain_8.8_coor}, \subref{Plar_realized_gain_10_coor}, and \subref{Plar_realized_gain_11.1_coor}, respectively. These plots further validate the simulations.

Upon examining the forward beams in Fig. \ref{Plar_realized_gain_HFSS_coor} or \subref{Plar_realized_gain_CST_coor}, it becomes evident that the realized gain and the half power beam width (HPBW) of the forward beams increase simultaneously with frequency. This phenomenon may appear confusing at first; however, a closer look at the 3D display of the patterns clarifies the situation. The increase in frequency causes the beam to focus more along the azimuthal angle (\(\phi\)), leading to a greater concentration of energy in the plane at \(\phi = 90^\circ\). As a result, both the realized gain and HPBW rise.

It's important to note that in such antennas, lower HPBW values observed in a 2D pattern do not necessarily indicate higher directivity or gain. This highlights the significance of presenting the radiation pattern in a 3D display, a detail that is often missing in some research papers. 

\section{conclusion}

The article discusses a spin PTI based on a kagome lattice, highlighting its unit cell and a three-dimensional dispersion diagram. The topological bandgap shows a 33$\%$ increase compared to 60-degree rhombic units, despite similar shapes. Edge modes are observed within the bandgap for a single ribbon, occurring without a gap in both slow and fast wave regions. The interface line, shaped like a Koch snowflake, demonstrates robustness and unidirectional propagation of pseudospins with sharp turns.

A parametric study optimized dimensions for the topological bandgap in the X-band frequency range, validated by the ribbon's dispersion diagram. The interface in an armchair configuration allows edge modes to be positioned in the fast wave region, ensuring effective excitation when coupled with the classical line. The S-parameters show effective coupling, and the realized gain pattern indicates scanning ranges of 50 and 47 degrees for backward and forward radiation, respectively. Table \ref{tableC} compares antennas based on different PTIs, with the proposed antenna achieving the widest scanning range and simultaneous backward and forward scanning.

The consistency of the simulation results from Ansys HFSS and CST Studio supports the certification of these results, as demonstrated in Fig. \ref{Kagome_hex_S_parameters_13} for the S-parameters and in Fig. \ref{Plar_realized_gain_8.8_coor} - \subref{Plar_realized_gain_11.1_coor} for the radiation patterns. Additionally, all of the simulation findings align with the dispersion diagram presented in Fig. \ref{Kagome_Hex_Ribbon}. Therefore, we can conclude that the simulation results are valid and reliable.

\section*{Acknowledgment}
We would like to thank Prof. A. Zeidaabadi Nezhad form Isfahan University of Technology and Dr. R.J.B. Davis from University of California, San Diego for the fruitful and illuminating discussions.

\bibliographystyle{IEEEtran}
\bibliography{references1}
\newpage
\begin{IEEEbiography}[{\includegraphics[width=1in,height=1.25in,clip,keepaspectratio]{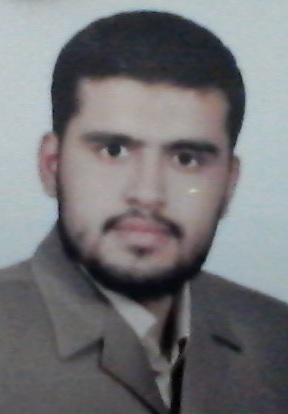}}]{SAYYED AHMAD ABTAHI} received the B.Sc. degree in electrical engineering from the Yazd University, Yazd, Iran, in 2014, and the M.Sc. degrees in telecommunications from Kashan University, Kashan, Iran, in 2018. He is currently
pursuing the Ph.D. degree in telecommunications with Isfahan University of Technology (IUT), Isfahan, Iran.  His research interests include topological metamaterials, leaky-wave antennas, periodic structures and photonic crystals.
\end{IEEEbiography}\hfil
\begin{IEEEbiography}[{\includegraphics[width=1in,height=1.25in,clip,keepaspectratio]{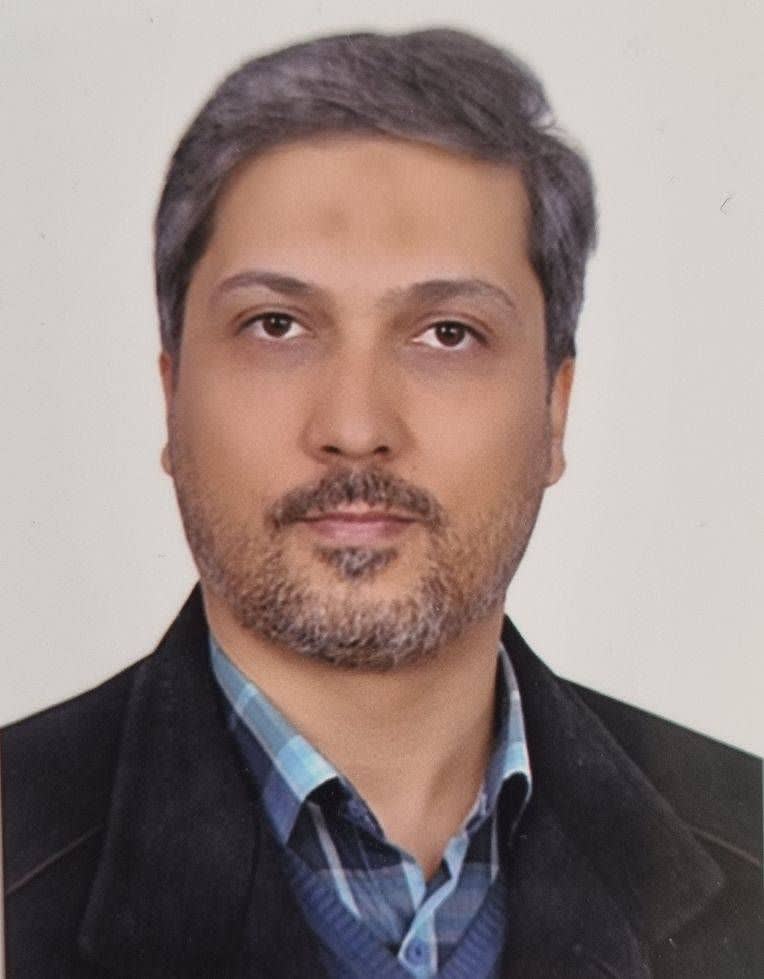}}]{MOHSEN  MADDAHALI} received the B.Sc. degree in electronics and telecommunications from the Isfahan University of Technology (IUT), Isfahan, Iran, in 2005, and the M.Sc. and Ph.D. degrees in electrical engineering from Tarbiat Modares University, Tehran, Iran, in 2008 and 2012, respectively.,Since 2012, he has been an Assistant Professor with the Department of Electrical and Computer Engineering, IUT. His research interests include computational electromagnetics, antenna theory, array antennas, new technologies in antenna design, and phased array antennas.
\end{IEEEbiography}
\begin{IEEEbiography}[{\includegraphics[width=1in,height=1.25in,clip,keepaspectratio]{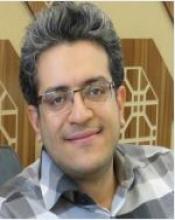}}]{AHMAD  BAKHTAFROUZ}received the B.S., M.S., and Ph.D. degrees from the Isfahan University of Technology (IUT), Isfahan, Iran, in 2006, 2009, and 2015, respectively, all in electrical engineering.,In 2016, he joined the faculty of the IUT, where he is currently an associate professor with the Electrical and Computer Engineering Department. His current research interests include millimeter-wave antennas, plasmonic devices, and periodic structures, such as electromagnetic bandgap structures (EBGs), artificial magnetic conductors (AMCs), and frequency selective surfaces (FSSs).
\end{IEEEbiography}

\EOD

\end{document}